\documentclass[ipo,apj]{emulateapj}
\usepackage{graphicx}
\usepackage{natbib}
\usepackage{amssymb}
\usepackage{amsmath}
\usepackage{float}
\usepackage{chngcntr}
\usepackage{fancyvrb}
\usepackage{xcolor}
\usepackage{soul}

\def\beq{\begin{equation}}
\def\eeq{\end{equation}}

\SaveVerb{mfm12}|fiducial_MFMlarge|
\SaveVerb{mfm08}|fiducial_MFMmid|
\SaveVerb{mfm04}|fiducial_MFMsmall|
\SaveVerb{fiducial}|fiducial|

\begin{document}

\title{Collisional Fragmentation Is Not a Barrier to Close-in Planet Formation}

\author{Joshua Wallace\altaffilmark{1}$^{,\star}$,
  Scott Tremaine\altaffilmark{2,1},
  and John Chambers\altaffilmark{3}
}
\affil{
  $^{1}$ Department of Astrophysical Sciences, Princeton University,
  Princeton, NJ 08544, USA \\
  $^{2}$ Institute for Advanced Study, Princeton, NJ 08540, USA \\
  $^{3}$ Department of Terrestrial Magnetism, Carnegie Institution 
  for Science, Washington, DC 20015, USA \\
}


\begin{abstract}
Collisional fragmentation is shown to not be a barrier to rocky planet
formation at small distances from the host star. Simple analytic arguments
demonstrate that rocky planet formation via 
collisions of homogeneous gravity-dominated bodies is 
possible down to distances of order the Roche radius
($r_\mathrm{Roche}$).  Extensive N-body simulations with initial
bodies $\gtrsim\!1700$ km that include
plausible models for fragmentation and merging of gravity-dominated
bodies confirm this conclusion and  demonstrate that rocky planet
formation is possible down to ${\sim}$1.1 $r_\mathrm{Roche}$.  At
smaller distances, tidal effects cause collisions to be too
fragmenting to allow mass build-up to a final, dynamically stable
planetary system. We argue that even differentiated bodies can
accumulate to form planets at distances that are not much larger than
$r_\mathrm{Roche}$. 

\end{abstract}


\slugcomment{Submitted to AAS Journals}


{\let\thefootnote\relax\footnote{$^{\star}$\href{mailto:joshuajw@princeton.edu}{joshuajw@princeton.edu}}}
\setcounter{footnote}{0}
\section{Introduction}

\label{sec:intro}

Exoplanetary systems possess a
large and surprising diversity of architectures. 
In particular, many 
systems contain one or more planets with periods
$\lesssim$10~days.
A partial list of examples includes 
Kepler--42, a
${\sim}0.13$ M$_\odot$ star with three known planets, all rocky, with
periods ${\sim}$0.5, ${\sim}$1.2, ${\sim}$1.9 days
(\citealt{muirhead2012}); 
Kepler--32,   a ${\sim}$0.58 M$_\odot$ star
with five known planets, 
having periods between ${\sim}0.7$ and ${\sim}22$ days 
(\citealt{muirhead2012}); and
TRAPPIST--1, a ${\sim}$0.08 M$_\odot$ star with  seven
known planets, all with radii ${\sim}1$~R$_\oplus$ and
 periods between ${\sim}$1.5 and ${\sim}$19 days
(\citealt{gillon2017}).

A central question is whether (i) such
close-in 
planetary systems can form {\it in situ}, or (ii) they formed further out from their host star and migrated inwards to
their current configuration. There are difficulties with both
mechanisms. (i) Assuming solar metallicity, the
gas surface density of a disk containing sufficient metals
for {\it in situ} formation of close-in planets such as those
highlighted above would be gravitationally unstable
(\citealp{raymond2014,schlichting2014}).  Such disks should therefore form
giant rather than small rocky planets.
Moreover, even if the disk were stable, the high temperatures
(${\sim}$2000 K) expected
in the inner disk would prevent 
the condensation of dust within 0.1 AU (\citealt{dalessio1998}).
A possible solution is that inward migration of solid material
through the protoplanetary disk could allow for sufficient
planet-building material to accrue at small semi-major axes 
without an unstable build up of gas
(e.g., \citealp{
youdin2002, youdin2004,chiang2010};
\citealp{chatterjee2014,chatterjee2015}). (ii) Although the
physics of planet migration is robust, the behavior
of migrating planets depends sensitively on the properties and physics
of protoplanetary disks \citep[e.g.,][]{baruteau+2014}. Simple models
of migration incorrectly predict that most short-period planets should
be in orbital resonances, and cannot explain the large numbers of
short-period planets discovered by the {\it Kepler} spacecraft \citep[e.g.,][]{benz+2014}.

In this paper, we examine  another possible barrier to {\it in situ}
formation of close-in planets: collisional fragmentation.
Even if planetesimals 
or planetary embryos can form at or migrate to these short-period
orbits, would the relative velocities between such bodies allow 
for merging and growth, or would they be
sufficiently large that collisions would be primarily fragmenting? 
In the latter case, planetesimals would not be able to grow into
planets. We address this question both analytically and with N-body
simulations. 

The destruction of bodies due to collisions has been studied in depth
in the context of asteroids, the Kuiper belt, and debris disks 
(e.g., \citealt{dohnanyi1965,wyatt2002, dominik2003, kobayashi2010,
kenyon2017a, pan2005}). 
However, the bodies focused on in these studies are 
$\lesssim\!100$~km in size (and many are $\lesssim\!1$~km), much
smaller than the bodies that will be 
examined in this work.    The physics of fragmentation is different in
small and large bodies, with small bodies ($\lesssim\!0.1$--1~km
for rocky bodies) being held
together primarily by their internal strength (and thus affected by
internal flaws and cracks) and large bodies being held together
primarily by their gravitational self-attraction (although many small bodies are actually rubble piles with little
internal strength; we note \citealt{pan2005} assumed this in their collision models). The results
of a fragmenting collision also depend on body size:
larger bodies have significant escape velocities and thus some of the
fragments from an impact can be reaccumulated. 
Collision
velocities between large bodies, especially late in the planet-formation process when they are on well-separated and nearly stable
orbits, can also differ from those of small bodies, which have random
velocities that are themselves often set by stirring from the
large bodies. 
In particular, the models of debris-disk production via
collisional cascades are generally based on a bimodal mass distribution with the
largest bodies having undergone runaway growth and stirring up the
smaller bodies to sufficiently high random velocities to have
destructive collisions.  In this picture, the larger remnants from
runaway growth are what eventually form the final planets and are the
focus of the present study.
For these reasons, results from collisional cascade models of small
bodies cannot be assumed to carry over to the large bodies that
eventually form planets.

In Section~\ref{sec:analytic} we provide an analytic motivation for
our study, then in Section~\ref{sec:method} we describe the methods
 of our N-body simulations.  We present the results from the
 simulations in Section~\ref{sec:results}, provide a discussion 
 in Section~\ref{sec:discussion}, and conclude in
 Section~\ref{sec:conclusion}.

\section{Analytic Motivation}
\label{sec:analytic}

\subsection{Preliminaries}
Here we present an order-of-magnitude argument describing 
the necessary conditions for collisions between gravity-dominated bodies to be
erosive.  We define ``erosive'' to mean that  the largest remnant
after a collision is less massive 
than the larger of the two colliding bodies.   Throughout this work, we
always assume that the colliding bodies have 
sufficient mass that their gravitational binding energy
is greater than their molecular binding energy (the gravity-dominated
regime).  We also assume that collisions are between two 
spherical bodies of the same density, having 
masses $m_1$ and $m_2$ and radii $R_1$ and $R_2$. We specify $m_1\!\geq\! m_2$ and, following common
convention, call the larger body
involved in the collision the ``target'' and the smaller body the
``projectile''.  In this section, we assume all orbits are
coplanar (no mutual inclination).

The amount of fragmentation that occurs in a collision depends on the
ratio between 
the specific impact energy in the center-of-mass frame $Q$ and
some threshold 
energy $Q^*$, defined to be the specific impact energy necessary to
disperse half the mass involved in the collision.  The impact energy
is 
\begin{equation}
\label{eq:q}
Q = \frac{\mu (\Delta v)^2}{2(m_1+m_2)},
\end{equation}
where $\mu$ is the reduced mass of the two colliding bodies and $\Delta v$
their collision velocity (the relative velocity just before
impact). From, e.g., \cite{stewart2012} 
\begin{equation}
\label{eq:qstar}
Q^* = \left(\frac{\mu}{\mu_\beta}\right)^{3/2}\left[\frac{(1+\alpha)^2}{4\alpha}\right]\frac{4}{5}C^*\pi\rho_1GR^2_{C1},
\end{equation}
where $\mu_\beta$ is the
reduced mass of the target body and the 
fraction $\beta$ of the projectile body that intersects the target body during
a collision, which for off-center collisions may be less than unity (see
\citealt{stewart2012} for details),  
$\alpha\! =\! m_2/m_1$, 
$C^*$ is a dimensionless factor that accounts for the dissipation of
energy in the target body, 
$\rho_1\!=\!1$ g 
cm$^{-3}$, $G$ is the gravitational constant, and $R_{C1}$ is the
radius of a body with density $\rho_1$ and 
a mass equal to the sum of the projectile and target masses,
\begin{equation}
\label{eq:rc1}
\frac{4\pi\rho_1}{3}R^3_{C1} = m_1 + m_2.
\end{equation} 
\cite{leinhardt2012} (hereafter LS12) found the best-fit value of
$C^*$ for planet-sized bodies is 
$1.9\!\pm\!0.3$; \cite{chambers2013} (hereafter C13) used
$C^*\!=\!1.8$ and we use that value in this work.
Note that the expressions of Equation~(\ref{eq:qstar})
in \cite{stewart2012} (primarily their 
equation 10) are more general because they contain an extra parameter,
the velocity exponent $\overline{\mu}$. Following C13 and consistent
with the best-fit values found by \cite{stewart2012}, we use
$\overline{\mu}\!=1/3\!$ to get Equation~(\ref{eq:qstar}). 

In this section, we assume that collisions are between bodies on
initially well-separated, 
near-circular orbits whose eccentricities have been gradually excited
by mutual perturbations. This situation arises, for example, in the
late stages of planet formation after oligarchic growth is complete
and the population of residual small bodies has decayed. (This is in
contrast to the early stages of planet formation, where there is a
much higher density of colliding bodies and the random velocities are
set by the escape velocity of the largest bodies.)
If the two bodies initially have semi-major axes $a \pm \Delta a/2$,
with $\Delta 
a \ll a$, then their
relative speed $(\Delta v)_{\rm rel}$---when they are close to
colliding but far enough away that gravitational focusing is
unimportant---will be
\begin{equation}
(\Delta v)_{\rm rel} = f_1 \left(\frac{GM_*}{a}\right) ^{1/2} \frac{\Delta a}{a} =
f_1 v_c \frac{\Delta a}{a},
\end{equation}
where $M_*$ is the mass of
the central star, $v_c\!=\!\sqrt{GM_*/a}$ 
is the Keplerian 
circular velocity at $a$,  and $f_1$ is a factor of
order unity.  Two such bodies are unable to collide unless the
  eccentricity $e\! \gtrsim\! \Delta a/a$.
 Including gravitational focusing, the collision
  velocity $\Delta v$ between these two bodies will be
\begin{equation}
\Delta v = [(\Delta v)_{\rm rel}^2 + v_e^2]^{1/2},
\end{equation}
where $v_e$ is the mutual escape velocity, 
which we define as
\begin{equation}
\label{eq:vesc}
v_e = \sqrt{2G\frac{m_1 + m_2}{R_1 + R_2}}.
\end{equation}

\subsection{Equal-mass Case}
Assuming the bodies have the same
mass $m$ and density $\rho$ (and thus the same radius $R$) and
  have a head-on collision so $\mu\!=\!\mu_\beta$, 
the specific impact energy scaled by $Q^*$ is
\begin{align}
\frac{Q}{Q^*}  =& \frac{\mu}{2(m_1+m_2)}\frac{(\Delta v)^2}{Q^*}\\
= & \frac{1}{8}\frac{(\Delta v)_{\rm rel}^2 +
    v_e^2}{Q^*}  \\
\label{eq:approx_vrel}
\simeq & \frac{1}{8}\left[\frac{5f_1^2M_*(\Delta
    a)^2}{4\pi\rho_1C^*a^3R_{C1}^2} + 
    \frac{v_e^2}{Q^*}\right]\\
\label{eq:collision_velocity}
= &\frac{1}{8}\left[\frac{5f_1^2M_*(\Delta a)^2}{2^{4/3}3^{2/3}\pi^{1/3}\rho_1^{1/3}C^*a^3m^{2/3}} +
    \frac{v_e^2}{Q^*}\right].
\end{align}
The quantity $v_e^2/Q^*$ can be simplified as follows.  First,
\begin{equation}
\label{eq:vescreduced}
\frac{v_e^2}{Q^*} = \frac{5(m_1+m_2)}{2 \pi \rho_1 C^* (R_1+R_2)R_{C1}^2}.
\end{equation}
 Eliminating $\rho_1$ using Equation~(\ref{eq:rc1}), we get
\begin{equation}
\label{eq:vescreduced2}
\frac{v_e^2}{Q^*} = \frac{10R_{C1}}{3 C^* (R_1+R_2)}.
\end{equation}
For arbitrary masses, 
\begin{align}
R_{C1}=&\left[\frac{3(m_1+m_2)}{4\pi\rho_1}\right]^{1/3}, \\
R_1+R_2=&\left(\frac{3}{4\pi\rho}\right)^{1/3}(m_1^{1/3}+m_2^{1/3}). 
\end{align}
Hence, for equal-mass bodies, 
\begin{align}
\frac{Q}{Q^*}  \simeq& \frac{5f_1^2M_*(\Delta
  a)^2}{2^{13/3}3^{2/3}\pi^{1/3}\rho_1^{1/3}C^*a^3m^{2/3}} 
\nonumber\\
\label{eq:pluginveoverq}
 &~~~~~+
\frac{5}{3C^*2^{8/3}}\left(\frac{\rho}{\rho_1}\right)^{1/3}.  
\end{align}

Equation~(\ref{eq:pluginveoverq}) can be rewritten in terms of the
mutual Hill radius, ${r_\mathrm{Hill}\!=\!a[(m_1+m_2)/(3M_*)]^{1/3}}$,
\begin{align}
\frac{Q}{Q^*} \simeq&
\frac{5f_1^2}{2^{11/3}3^{4/3}\pi^{1/3}C^*}\left(\frac{M_*}{\rho_1
    a^3}\right)^{1/3}\frac{(\Delta a)^2}{r^2_\mathrm{Hill}} 
\nonumber \\
\label{eq:expressed_in_Hill}
&~~~~~~~~~~~~~~~+
\frac{5}{3C^*2^{8/3}}\left(\frac{\rho}{\rho_1}\right)^{1/3}.
\end{align}
The criterion that head-on collisions of equal-mass bodies are
  non-erosive is $Q\!<\!Q^*$. For later use we shall write this as
  $Q\!<\!f_2Q*$ with $f_2\!=\! 1$ for equal-mass bodies.
Thus we obtain a necessary condition for
accumulation, 
\begin{align}
& \frac{5f_1^2}{2^{11/3}3^{4/3}\pi^{1/3}C^*}\left(\frac{M_*}{\rho_1
    a^3}\right)^{1/3}\frac{(\Delta a)^2}{r^2_\mathrm{Hill}}
\nonumber \\
\label{eq:condition}
&~~~~~~~\lesssim \,f_2
-\frac{5}{3C^* 2^{8/3}}\left(\frac{\rho}{\rho_1}\right)^{1/3}.
\end{align}
Equation~(\ref{eq:condition}) sets a maximum value for the initial
separation $\Delta a$ at
which two colliding bodies avoid an erosive collision. 
For future use, we note that the relation between $f_2$ and the
minimum collision 
velocity for erosive head-on collisions  between equal-mass
  bodies can be written
\begin{equation}
\frac{(\Delta
  v)^2}{v_e^2}=\frac{3\cdot2^{8/3}}{5}f_2C^*\left(\frac{\rho_1}{\rho}\right)^{1/3}.
\label{eq:asphaug}
\end{equation}

If the typical separation in the post-collision
system (which we call $\Delta a'$) is too small, then the system will
be dynamically unstable and eventually undergo another
  collision.  An approximate stability criterion is  
$\Delta a'\!\gtrsim\! f_3 r'_\mathrm{Hill}$, where $f_3\!\simeq\! 10$
for stability over  
$10^{10}$ orbits (\citealt{pu2015})
and the primes are used to distinguish post-collision properties from
pre-collision properties. We note the chaotic nature of
  collisional growth, but still choose to employ this criterion as it is a
  necessary one for 
  stability in the long-lived post-collision stage, not the short-lived
  chaotic collision stage.  This gives us
 a lower bound on dynamically
stable $\Delta a'$, while   Equation~(\ref{eq:condition}) gives us an
upper bound on $\Delta a$ for collisions to be non-erosive.
If we assume a system of bodies with equal masses $m$,
separated  from their nearest companions in semi-major axis by
$\Delta a$ just prior to their last collision before stability, and if
we also assume that adjacent bodies collide and merge 
pairwise (this ignores the possibility of
higher-eccentricity collisions between non-adjacent bodies),  then 
 the post-collision bodies will be separated from each other by
$\Delta a'\simeq 2\Delta a$ and will have 
$r'_\mathrm{Hill}\! \simeq\! 2^{1/3}r_\mathrm{Hill}$.
  Thus, the post-collision stability
condition $\Delta a'\!\gtrsim\! f_3 r'_\mathrm{Hill}$ can be rewritten
in terms of 
pre-collision quantities, $2^{2/3}\Delta a \!\gtrsim\! f_3 r_\mathrm{Hill}$.
Plugging this into Equation~(\ref{eq:condition}) to eliminate $\Delta
a$, 
we obtain a range for $a$ in which collisions will not be erosive up
to the final collision before long-term dynamical stability,
\begin{align}
a \gtrsim \,&
\frac{5f_1^2f_3^2}{32\pi^{1/3}3^{4/3}C^*}\left(\frac{M_*}{\rho_1
    }\right)^{1/3}
\nonumber \\
\label{eq:final_condition}
& ~~~~~\times\left[f_2-\frac{5}{3C^*2^{8/3}}\left(\frac{\rho}{\rho_1}\right)^{1/3}\right]^{-1}.
\end{align}
Remarkably, this criterion is independent of the planetesimal mass $m$.

The Roche radius $r_\mathrm{Roche}$ for a homogeneous body
of density $\rho$ is
\begin{align}
r_\mathrm{Roche} &= 1.523\left(\frac{M_*}{\rho}\right)^{1/3}
\nonumber \\ 
 &= 0.0089 ~\mathrm{AU} \left(\frac{M_*}{M_\odot}\frac{3~\textrm{g
    cm}^{-3}}{\rho}\right)^{1/3}.\label{eq:roche}   
\end{align}
If a gravity-dominated body on a circular orbit has $a\! <\!
r_\mathrm{Roche}$, it is disrupted by tidal forces.
Writing Equation~(\ref{eq:final_condition}) in terms of
$r_\mathrm{Roche}$ and using $C^*\!=\!1.8$, we get
\begin{align}
a \gtrsim \,& \frac{5f_1^2f_3^2}{32\pi^{1/3}3^{4/3}C^*}\left(\frac{\rho}{\rho_1
    }\right)^{1/3}\frac{r_\mathrm{Roche}}{1.523} \nonumber \\
&
~~~\times\left[f_2-\frac{5}{3C^*2^{8/3}}\left(\frac{\rho}{\rho_1}\right)^{1/3}\right]^{-1}
\nonumber \\
\label{eq:intermsofRroche}
\gtrsim \,& 0.013~ \frac{f_1^2f_3^2\eta^{1/3}}{f_2-0.21\eta^{1/3}}
r_\mathrm{Roche}, 
\end{align}
where $\eta = \rho / 3$ g cm$^{-3}$ and $\rho/\rho_1\!=\!3\eta$.
This is the semi-major axis range in which the last collisions before
long-term dynamical stability will not be
erosive. 
Apart from the dimensionless factors $f_1,f_2,f_3$, the value of $a$ at which a system
transitions from non-erosive to erosive collisions is less than
$r_\mathrm{Roche}$.\footnote{One caveat is that we do not consider the
effects of the strong tides present near $r_\mathrm{Roche}$, which may
reduce $Q^*$. The effect of tides is discussed further in Section \ref{sec:3.1}.}  Since gravity-dominated bodies cannot survive
within $r_\mathrm{Roche}$, to decide whether collisional fragmentation
is an additional barrier to growth we must determine whether the
transition point is inside or outside $r_\mathrm{Roche}$.

We now determine reasonable values for $f_1$, $f_2$, and $f_3$.  As
noted previously, 
\cite{pu2015} found that $f_3\!\simeq\!10$. A linear expansion in
  $a$ of 
orbital velocity $v$ around $v_c$ gives $f_1\! \simeq\! 0.5$. 
As mentioned earlier, for
  equal-mass bodies, $f_2\!=\!1$.
If we scale 
$f_1$, $f_2$, and $f_3$ by these values, 
then from Equation~(\ref{eq:intermsofRroche}) we find
\begin{align}
\label{eq:combined}
a \gtrsim a_{\textrm{frag}}  = & 0.41 ~r_\mathrm{Roche}\nonumber \\
&\quad\times \left(\frac{f_1}{0.5}\right)^2\left(\frac{f_3}{10}\right)^2\eta^{1/3}~
\frac{
  1 -0.21}{f_2-0.21\eta^{1/3}}\\ 
  = & 0.0036
  ~\mathrm{AU} \left(\frac{M_*}{M_\odot}\right)^{1/3}\nonumber \\ 
 &~~~~~\times\left(\frac{f_1}{0.5}\right)^2\left(\frac{f_3}{10}\right)^2 \frac{
  1 -0.21}{f_2-0.21\eta^{1/3}}, 
\end{align}
where $a_{\textrm{frag}}$ is the semi-major axis at which collisions
between equal-mass bodies
transition from non-erosive to erosive.
For comparison, the solar radius is $0.0047$ AU.  
For these values of $f_1$, $f_2$, and $f_3$, the semi-major axis range in
which collisions are primarily erosive will {\it
  always} be less than $r_\mathrm{Roche}$, i.e., the bodies will
be tidally disrupted before they are close enough to their host star
to reach high enough collision velocities to erode one another. 
 If we have chosen correct values for $f_1$,
$f_2$, and $f_3$, then Equation~(\ref{eq:combined}) says that
gravity-dominated bodies, once formed, are able to 
engage in non-erosive collisions 
for semi-major axes all the way down to
$r_\mathrm{Roche}$, independent of 
stellar mass and the mass and density of the bodies.   However, 
modifications to $f_1$, $f_2$, and $f_3$ could lead to erosive collisions
outside $r_\mathrm{Roche}$. We give two brief examples:

\begin{itemize}

\item \cite{kenyon2017a} model collisional cascades in
  strength-dominated bodies and find that
$(\Delta v)^2/Q^*\! \geq\! 5$ led to erosion of the largest
bodies. Equation (\ref{eq:q}) implies that $Q\!<\! (\Delta v)^2/8$ for the
largest body in a continuous mass distribution, so this would mean
that $f_2\!\lesssim\!0.6$. A value $f_2=0.6$, with the fiducial 
parameters in Equation (\ref{eq:combined}), leads to 
$a_\mathrm{frag}\!=\!0.83$ $r_\mathrm{Roche}$.

\item Our results only hold for homogeneous bodies. Using
hydrodynamical simulations of collisions between differentiated 
bodies, \cite{asphaug2010} found that equal-mass head-on
collisions were non-erosive if $\Delta v \! < \! k v_e$ and $k\!
\simeq\! 2.9$. For our fiducial parameters $C^*\!=\!1.8$ and
$\rho/\rho_1\!=\!3$, Equation (\ref{eq:asphaug}) implies that
$k^2\!=\!4.75f_2$, so this value of $k$ implies $f_2\!=\!1.8$. 
However,
Asphaug also found that the crusts and
mantles of differentiated bodies are more easily stripped in
collisions (particularly off-center collisions) than the outer parts
of homogeneous bodies.  His simulations show
that collisions with an angle $\theta$ between the relative velocity vector
and relative center-of-mass vector ${\sim} 30^\circ$ have 
$k\!\approx\! 1.5$ ($f_2\!=\!0.5$ for our fiducial parameters) and collisions with $\theta\!\gtrsim\!45^\circ$
have even lower values, $k\!\approx\! 
1.2\text{--}1.4$ ($f_2\!=\!0.3$--0.4). If $f_2\!=\!0.3$--0.4 and $f_1$,
$f_3$, and $\eta$
respectively remain 0.5, 10, and 1, then from
Equation~(\ref{eq:combined}), 
$a_{\textrm{frag}}/r_\mathrm{Roche}=3.6$--1.7. 
However, the fractional mass loss in these off-center
collisions 
remains quite low over a large range of $k$.  For
$\theta\!=\!30^\circ$  the fractional mass loss is $\lesssim\!5$\% for
$k\!\lesssim\!1.7$ ($f_2\!\simeq\!0.61$), which occurs for
$a\! >\! 0.81$ 
$r_\mathrm{Roche}$.  Additionally, the mass loss for a given $k$ decreases with
increasing $\theta$. Equal-mass collisions with $\theta\!=\!45^\circ$ have
$\lesssim\!5$\% fractional mass loss for $k\!\lesssim\!2.2$
($f_2\!\simeq\! 1.1$), and $\theta\!=\!60^\circ$ gives ${\sim}2$\%
fractional mass loss even for $k\!=\!3$ ($f_2\!=\!1.9$).
The shallow slope of mass loss versus 
collision velocity for off-center collisions is balanced by the mass
growth from low-velocity off-center and low- and high-velocity head-on
collisions. These arguments suggest that our conclusion for
homogeneous equal-mass bodies, that 
collisional fragmentation is not a barrier to planet formation all the
way down to $r_\mathrm{Roche}$, likely holds for differentiated
bodies as well.

\end{itemize}

\subsection{Unequal-mass Case}
\label{subsec:unequalmass}
The preceding arguments can be generalized to bodies with unequal
  masses, though 
not quite as cleanly.
Assuming that the collision is sufficiently close to head-on
  that $\mu\!=\!\mu_\beta$ (and with $m_1\! \geq\! m_2$ and 
  $\alpha\! =\! m_2/m_1$ as before), we can
write a more general form of Equation~(\ref{eq:collision_velocity}),
\begin{align}
\frac{Q}{Q^*}  =& \frac{\mu}{2(m_1+m_2)}\frac{(\Delta v)_{\rm rel}^2 +
    v_e^2}{Q^*}\\
=&\frac{\alpha}{2(1+\alpha)^2}\left[\frac{5\alpha f_1^2M_*(\Delta a)^2}{(1+\alpha)^2\pi\rho_1C^*a^3R_{C1}^2} +
    \frac{v_e^2}{Q^*}\right] \\
\label{eq:collision_velocity_modified}
= &\frac{\alpha}{2(1+\alpha)^2}\nonumber \\
\times&\left[\frac{5\cdot 2^{4/3}\alpha f_1^2M_*(\Delta a)^2}{3^{2/3}(1+\alpha)^{8/3}\pi^{1/3}\rho_1^{1/3}C^*a^3m_1^{2/3}} +
    \frac{v_e^2}{Q^*}\right].
\end{align}
 Still assuming that both bodies have the same
density, we may follow steps similar to those in Equations
(\ref{eq:vescreduced})--(\ref{eq:pluginveoverq}) to obtain
\begin{align}
\frac{Q}{Q^*} =& \frac{\alpha}{2(1+\alpha)^2}\Biggl[
\frac{5\cdot 2^{4/3}\alpha f_1^2M_*(\Delta a)^2}{3^{2/3}(1+\alpha)^{8/3}\pi^{1/3}\rho_1^{1/3}C^*a^3m_1^{2/3}}
%
%
\nonumber \\ 
&~~~+\frac{40}{3C^*}\frac{\alpha}{(1+\alpha)^{5/3}(1+\alpha^{1/3})}
    \left(\frac{\rho}{\rho_1}\right)^{1/3}\Biggr].
\label{eq:collision_velocity_modified_a}
\end{align}
We substitute $r_\mathrm{Hill}$ as defined after
Equation~(\ref{eq:pluginveoverq}), now expressed in terms of $m_1$
and $\alpha$, into 
Equation~(\ref{eq:collision_velocity_modified_a}) and obtain more general
versions of Equation~(\ref{eq:expressed_in_Hill}),
\begin{align}
\frac{Q}{Q^*} =&
\frac{\alpha^2}{(1+\alpha)^4} \nonumber \\
&\times\Biggl[\frac{5\cdot
  2^{1/3}f_1^2}{3^{4/3}\pi^{1/3}C^*}\left(\frac{M_*}{\rho_1
  a^3}\right)^{1/3}\frac{(\Delta a)^2}{r^2_\mathrm{Hill}}\nonumber \\
 &~~~+   \frac{20}{3C^*}\frac{(1+\alpha)^{1/3}}{1+\alpha^{1/3}}
    \left(\frac{\rho}{\rho_1}\right)^{1/3}\Biggr], 
\end{align}
and Equation~(\ref{eq:condition}),
\begin{align}
& \frac{(1+\alpha)^4}{\alpha^2}f_2 - \frac{20}{3C^*}\frac{(1+\alpha)^{1/3}}{1+\alpha^{1/3}}
    \left(\frac{\rho}{\rho_1}\right)^{1/3}  \nonumber \\
&~~~~~~\gtrsim \frac{5\cdot
  2^{1/3}f_1^2}{3^{4/3}\pi^{1/3}C^*}\left(\frac{M_*}{\rho_1
  a^3}\right)^{1/3}\frac{(\Delta a)^2}{r^2_\mathrm{Hill}}
.
\end{align}
We use similar arguments as before regarding the pre- and
post-collision values for $\Delta a$ and $r_\mathrm{Hill}$, except
that we
relax the assumption that the masses of the bodies are all the same.  As before,
$\Delta a'\! \simeq\! 2 \Delta a$.  
  However, the post-collision mutual Hill radius $r_\mathrm{Hill}'$
   depends not only 
  on the mass of the newly 
  merged body  but also the masses of the nearest neighbors, which,
  in principle, could take any value.  Using $c$ to 
  express the mass ratio between the newly merged body (of mass
  $m_1 + m_2$) and the mass $m_3$ of a nearby body of interest, $m_3 =
  c(m_1+m_2)$, we 
  can write $r_\mathrm{Hill}'\! \simeq\! (1+c)^{1/3}r_\mathrm{Hill}$.
  Thus, the condition $\Delta 
  a'\!\gtrsim\! f_3 r_\mathrm{Hill}'$ becomes $2\Delta a\!\gtrsim\!(1+c)^{1/3}
  f_3 r_\mathrm{Hill}$. 
From this, we  obtain an analog to Equation~(\ref{eq:final_condition}) 
in the case of arbitrary masses,
\begin{align}
  a \,\gtrsim \,& \frac{5
  f_1^2f_3^2(1+c)^{2/3}}{2^{5/3}3^{4/3}\pi^{1/3}C^*}\left(\frac{M_*}{\rho_1}\right)^{1/3} \nonumber \\
 & \times \left[\frac{(1+\alpha)^4}{\alpha^2}f_2 - \frac{20}{3C^*}\frac{(1+\alpha)^{1/3}}{1+\alpha^{1/3}}
     \left(\frac{\rho}{\rho_1}\right)^{1/3}\right]^{-1}.
\end{align}
Expressing the result in units of $r_\mathrm{Roche}$, we find
\begin{align}
a\, \gtrsim\, & \frac{5
  f_1^2f_3^2(1+c)^{2/3}\eta^{1/3}}{2^{5/3}3\pi^{1/3}C^*}\frac{r_\mathrm{Roche}}{1.523}
\nonumber \\
\label{eq:arange_k}
 & \times \left[\frac{(1+\alpha)^4}{\alpha^2}f_2 - \frac{20\eta^{1/3}}{3^{2/3}C^*}\frac{(1+\alpha)^{1/3}}{1+\alpha^{1/3}}
     \right]^{-1}.
\end{align}

We now find the value of $\alpha$ that maximizes the right
side of Equation~(\ref{eq:arange_k}).  The parameter $f_2$ is a
function of $\alpha$ (see the first line of Equation 8 in C13) and is
expressed as such in our maximization. 
The value of $\alpha$ that maximizes  Equation~(\ref{eq:arange_k}), as
well as the maximum value, depend on the values of the various
parameters in  Equation~(\ref{eq:arange_k}). 
 The factor $f_1$ does not
depend on the masses of the bodies, and thus
will keep its same value.
We note that
most of the work on orbital stability has
focused on equal-mass bodies, so it is  not known whether
$f_3$ depends on relative masses\footnote{The dependence on the
  relative mass is likely to be weak, 
since Hill's equations describing the interactions of two small bodies
orbiting a much larger third body depend only on the sum of the masses
of the two small bodies (e.g., \citealt{henon1986}).}; we assume $f_3$
does not depend on 
mass.
Thus, for the maximization, we
assume as before $f_1\!=\!0.5$,  $f_3\!=\!10$,  $\eta\!=\!1$,  and
$C^*\!=\!1.8$, and 
we also assume  $c\!=\!1$. 
Since we have assumed $m_1\!\geq\!m_2$,
$\alpha$ must be in the interval $(0,1]$. The local maximum in
this domain using the assumed values of the other parameters is found
at $\alpha\!=\!0.49$ and  gives
\begin{equation}
\label{eq:rrocheunequalmass}
a \gtrsim a_{\textrm{frag,}\alpha} =  0.51 ~r_\mathrm{Roche},\\ 
\end{equation}
where $ a_{\textrm{frag,}\alpha}$ is the same as $a_\mathrm{frag}$ but
generalized to arbitrary $\alpha$.
Thus our
conclusions from the homogeneous equal-mass case should remain
approximately valid for 
arbitrary mass ratios. For differentiated
bodies, the arguments presented before based on the results
of \cite{asphaug2010} still hold, as for head-on collisions,
$f_2\!\simeq\!1.3$ for 
$\alpha\!=\!0.5$ and $f_2\!\simeq\!1.5$ for
$\alpha\!=\!0.1$ (which are the only two unequal mass ratios examined
in \citealt{asphaug2010}), and for off-center collisions, all $f_2$
values are greater for $\alpha\!=\!0.5,0.1$ collisions than the $\alpha\!=\!1$
collisions with corresponding $\theta$.

Our statement that collisions lead to growth if $Q\!<\!f_2Q^*$
ignores any dependence of $f_2$  
on stellar tides.  Close to $r_\mathrm{Roche}$, tides from the star
will make it easier for fragments to escape the gravitational
influence of whatever coalesced body may be left after a collision, 
and thus $f_2$ 
will take a smaller value than  at larger semi-major
axes.   This is accounted for in our numerical work by modifying the
escape speed, as described in
Section~\ref{sec:3.1}. An additional complication not addressed
  by our analytic work is that collisions
will take on a variety of impact parameters and have different
possible collision outcomes than  the pure fragmentation outcome
considered in this section.
Because of these additional considerations, and because of the
sensitivity of our analytic result to even small uncertainties 
in $f_1$, $f_2$, and $f_3$,  
it is essential to perform N-body simulations of 
systems of rocky bodies to test the analytic results of this section.  The
remainder of this work presents the methods and results of such
calculations. 


\section{Numerical Methods}
\label{sec:method}
Our approach is very similar to that of
C13, who studied the effect of collisional fragmentation on the
formation of rocky planets in a solar system context.  
Our simulations were
carried out using the hybrid-symplectic
integrator in the {\it Mercury} N-body integrator package
(\citealt{chambers1999}), including the modifications introduced in
C13 to describe 
collisional fragmentation.   We first describe the
collision algorithm and then discuss the initial conditions of our
simulations.

\subsection{Collision Algorithm}
\label{sec:3.1}
The prescription used for collision outcomes is that of  LS12 as  implemented in C13, with
modifications described  
below to account for stellar tides.  
Here we briefly describe the possible outcomes of a
collision and direct the reader to C13 for a full description of the algorithm. 
When a collision is identified, 
the relative velocity
(including gravitational focusing) at the time of
impact is calculated.  If the impact velocity is less than a
modified mutual escape velocity $v_e'$, then the two bodies merge. 
This modified 
mutual escape velocity takes into account the
relatively small Hill spheres of the bodies at the small semi-major
axes present in our simulations.  Instead of fragments needing to
escape to infinity to become unbound from the bodies, they need only
escape to the edge of the Hill sphere. This is accounted for by defining $v_e'$
as follows, 
\begin{equation}
\label{eq:veschillmod}
v_e'^2 = \max\left\{v_e^2 \left( 1 - \frac{R_t + R_p}{r_\mathrm{Hill}}\right),0\right\},
\end{equation}
where $R_t$ and $R_p$ are the radii of the target and the projectile,
$v_e$ is the escape speed defined in Equation~(\ref{eq:vesc}), 
and $r_\mathrm{Hill}$ is the mutual Hill radius defined
after Equation~(\ref{eq:pluginveoverq}). 
If the impact
velocity is greater than $v_e'$, then fragmentation occurs, and the 
largest remnant mass is calculated as outlined in Equation (8) of C13.

The mass from the impacting bodies not included in the largest
remnant is then divided into one or more equal-mass
fragments, with masses as close as possible to but not below a minimum
fragment mass (MFM) that is specified for each simulation.  If the remaining mass
is less than the specified MFM, then the collision
is treated as a merger and the total mass is placed in a single body.

The details of how non-head-on collisions are handled 
can be found in C13 and LS12.  High impact-parameter collisions have
other possible outcomes in addition to the merger--fragmentation scenario presented
above for head-on collisions.  These are:
\begin{itemize}
\item hit-and-run collisions, where the target
remains intact and the projectile is disrupted, in some cases adding mass to
the target;
\item graze-and-merge collisions, where an initial off-center collision
saps sufficient kinetic energy from the system for an eventual
merger.
\end{itemize}
We use the boundary identified by \cite{genda2012} between the
hit-and-run and graze-and-merge regimes, replacing the escape
velocity $v_{\mathrm{esc}}$ in their Equation~(16) with our modified
escape velocity $v_e'$ to account for stellar tides.

 As was mentioned before, The amount 
of mass in a collision that is dispersed by fragmentation, if 
any, depends on the ratio between $Q$ (given by Equation~\ref{eq:q}) and
$Q^*$ (given by Equation~\ref{eq:qstar}). \cite{stewart2009}
found that, in the catastrophic 
disruption regime, the largest remnant mass (and thus the mass
dispersed by fragmentation) is linear in this ratio.  However, this simple
relation appears not to hold in regions close to the
Roche radius (e.g., \citealp{karjalainen2007, hyodo2014}; see
also \citealt{kenyon2017b} for an application to collisional
cascades). Instead, 
the mass of the largest remnant 
depends on the shapes of the colliding bodies and the relative
magnitudes of the azimuthal, radial, and vertical components of the
impact velocity (for the radial velocity component in particular, the
dependence is not monotonic on specific impact energy). Since this
study is focused more on answering the question of whether Keplerian
orbit-driven collision velocities are ever large enough to prevent
planet formation and not on precisely how tidal effects affect
collisional planet formation, we approximate the tidal effects 
by replacing $Q^*$ with 
$Q^*_\textrm{tidal}$, defined as
\begin{equation}
\label{eq:qstartidal}
Q^*_\textrm{tidal}(r) =\left\{
                      \begin{array}{ll}
                       Q^* \left[1
                      - \left(r_\mathrm{Roche}/r\right)^3\right]
                      & \quad
                      r > r_\mathrm{Roche} \\
                      \\
                      0 & \quad r \leq r_\mathrm{Roche},
                      \end{array}\right.          
\end{equation}
where $r$ is the distance between the central star and the center of
mass of the colliding bodies. For
$r\!>\!r_\mathrm{Roche}$, we found that 
Equation~(\ref{eq:qstartidal}) predicts collision energies required to
nearly entirely disrupt bodies that are broadly consistent
with the simulations of \cite{hyodo2014}. Equation~(\ref{eq:qstartidal})
also converges to  the LS12 results far away from
the Roche radius, a result also found by \cite{hyodo2014}.
In the case $Q^*_\textrm{tidal}\! =\!
0$, any collision will cause the bodies to
fragment into as many fragments as possible, as limited by the MFM.  
Thus, collisions that occur
inside $r_\mathrm{Roche}$ will lead to complete disruption of the
colliding bodies. 
In our algorithm, disruption 
inside the Roche radius only occurs during a collision, i.e., if a
body migrates in to $r\!<\!r_\mathrm{Roche}$, it will not be
disrupted unless it suffers a collision while inside $r_\mathrm{Roche}$.

We also note that we are using the classical Roche
radius, which was derived for a strengthless, co-rotating body on a
circular orbit and that more general forms of the
disruption radius have been developed
(e.g., \citealp{holsapple2006,holsapple2008}).  However, 
the strengthless model approximates
collisional fragmentation well, since after a collision the fragments
are no longer bound to each other by tensile forces, only
gravity.  For this reason, and for simplicity, we use the
classical Roche radius in Equation~(\ref{eq:qstartidal}) and 
throughout the paper. We believe that the qualitative picture our results
paint is correct while quantitative details that depend on the Roche
radius may well be inaccurate. 

The choice of MFM for our collision algorithm represents a compromise
between computational cost and realism: if the MFM
is too large, a large fraction of fragmenting collisions will be
erroneously classified as mergers; if the MFM is too small,
then fragmentation  will
create so many fragments that the numerical calculations will slow to
a crawl.
To check that our results are not strongly dependent on the choice of MFM,
each set of initial conditions was run with three different values of the
MFM, as detailed in
Section~\ref{sec:initialconditions}.   

If multiple collisions involving the same body occur in the same time
step, only one of the collisions (chosen at 
random) was considered to have happened.  For a typical run, 
this situation happened at
most a few times over the course of the run.  Any bodies that pass inside the
solar radius (${\sim}$0.0047 AU) 
were assumed to merge with the central star.  Over the course of a typical
run, this happened only a few times.  Virtually all
such events involved bodies with small
masses (very near
the MFM), and the resulting total mass loss  was at most a few
percent of the total mass of the initial bodies.

\subsection{Initial Conditions}
\label{sec:initialconditions}

We carried out several sets of N-body integrations.  We first
describe the initial conditions of our fiducial set of simulations
then describe the variations we made to 
this fiducial set.  All of our systems orbit
a 1-M$_\odot$ star and we assume a fixed 
density of 3 g cm$^{-3}$ for all bodies, independent of mass.
A time step of $6\!\times\!10^{-3}$ days was used. 
 For simplicity, we 
assume that the star has a radius of 1 R$_\odot$, but note that the 
young stars in systems such as those we simulate may have considerably
larger radii (e.g., calculations by \citealt{baraffe+2015} 
show that a 1-M$_\odot$ star with an age of 3 Myr has a radius of $1.7R_\odot$).
Our 15 fiducial simulations  
start with 150 equal-mass bodies of 
individual mass 0.02 M$_\oplus$ (${\sim}1.6$  
times the mass of the Moon), corresponding to a radius of  ${\sim}2100$ km with the 3 g
cm$^{-3}$ density used in this work).  
This mass is roughly what is expected for bodies that form
during the runaway growth phase of planet formation
(e.g., \citealt{wetherill1989}) and thus our initial conditions can be
thought of as the beginning of the  oligarchic growth or
giant-impact stage of
planet formation. 

We ran our simulations assuming no gas was present, i.e., we assumed that the
protoplanetary gas disk at these radii had entirely disappeared by the beginning of our
simulations.  Thus our initial
conditions can be thought of as corresponding to
planetesimals/planetary embryos that were prevented from evolving into
orbit-crossing trajectories by eccentricity damping  while the gas 
was present, or planetesimals/planetary embryos that underwent
disk-driven migration and were frozen in semi-major axis 
as the disk evaporated away.
We caution that this assumption may not be realistic in some
or perhaps all disks.  The existence of planets like GJ 1214 b
(\citealt{charbonneau2009})---a ${\sim}7$ M$_\oplus$ planet at a
period of 1.58 days that is believed to have a substantial gas
envelope---implies that in at least some cases the formation of planets
with ${\sim}1$-day 
orbital periods has occurred before the gas has completely disappeared.

The initial semi-major axes of the bodies
were distributed between 0.005 and 0.04 AU using a similar
disk surface-density profile $\sigma(a)$ as \cite{chambers2001} and
C13: $\sigma 
\!\propto\! a^{-3/2}$ from 0.02~AU to 0.04~AU and, inside 0.02~AU,
$\sigma$ linear in $a$, 
starting  from $\sigma\!=\!0$ at 0.005~AU and increasing to a value
matching the  $a^{-3/2}$ 
profile at 0.02~AU. A cumulative distribution function was calculated
from this disk profile and its inverse was sampled in 150 equally
spaced locations to determine the initial semi-major 
axes of the bodies. This 
semi-major axis range was chosen to 
straddle the Roche radius (0.0089 AU for a 1-M$_\odot$ star with
planet density $\rho\!=\!3$ g cm$^{-3}$), with the lower bound 
chosen to be close to the solar radius (0.0047 AU).  We note that one
would not expect large rocky bodies to be found inside
$r_\mathrm{Roche}$, so it is unphysical to include such
bodies in our initial conditions but doing so is useful
 to verify our analytic results.
Initial eccentricities and inclinations were drawn
from a Rayleigh distribution, $f(x)\! =\! x /\sigma^2
\exp{[-x^2/(2\sigma^2)]}$, 
with $\sigma\!=\!0.01$ for eccentricity and  $\sigma\! =\!
0^{\circ}.5$ for inclination. 
The initial arguments of periapsis, ascending nodes, and mean
anomalies were chosen at random from uniform distributions in the
interval $[0,2\pi)$.  

As
described in Section~\ref{sec:3.1},
the MFM sets a floor to the size of fragments in our 
simulations.  
To investigate how the choice of MFM affects
our results, we employed three different
MFMs, which are 5, 7.5, and 15
times smaller than the initial masses of the bodies used in our
fiducial simulations, or, respectively, $4.0\!\times\!10^{-3}$,
${\sim}2.67\!\times\!10^{-3}$, and 
${\sim}1.33\!\times\!10^{-3}$ M$_\oplus$.  At the density used in this
work (3 g cm$^{-3}$), these 
bodies have radii of ${\sim}$1200, ${\sim}$1100, and ${\sim}$900
km. We ran five 
simulations for each combination of initial conditions and MFM. 
A smaller MFM means 
that more fragments, on average, are produced per collision.  Since
the time 
required to calculate gravitational forces goes as
the square of the number of bodies, increasing the number of fragments
by decreasing the MFM 
 significantly slows down the calculation.  This slowdown is
exacerbated by the accumulation of fragments inside $r_\mathrm{Roche}$, 
which are unable to merge and thus persist for the duration
of the simulation. Therefore the choice for the smallest MFM
was limited by computation time.  The simulations were run for
$3\!\times\!10^5$ years with the exception of the runs with the
smallest MFM, 
${\sim}1.33\!\times\!10^{-3}$ M$_\oplus$, which lasted for
only $1\!\times\!10^5$ years, due to limits on computation time.

We adopted a naming convention to organize our numerical results.  A name
defines a ``set of simulations'', by which we mean simulations that have identical 
initial conditions and MFM but use different seed values 
for the random number generator.  The initial
orbital parameters that are chosen at random are eccentricity,
inclination, argument of
periapsis, ascending node, and mean anomaly. Thus, simulations
in a set share identical initial body masses, numbers of bodies, 
semi-major axes, and MFM.
Our three fiducial sets of simulations 
are described above.  The fiducial set 
with 
$\mathrm{MFM}\!=\! 4.0\!\times\!10^{-3}$ M$_\oplus$ is named
\verb!fiducial_MFMlarge!, 
the fiducial set with 
$\mathrm{MFM}\!\simeq\! 2.67\!\times\!10^{-3}$ M$_\oplus$ is named
\verb!fiducial_MFMmid!, and the fiducial set 
with $\mathrm{MFM}\!\simeq\! 1.33\!\times\!10^{-3}$ M$_\oplus$ is named
\verb!fiducial_MFMsmall!.  Each set in this work consists of
five individual simulations.

\begin{deluxetable}{p{0.65in}p{2.35in}}
\tablecaption{Naming Convention for Simulations\label{tab:run_names_table}}
\tablehead{\colhead{prefix}   &
           \colhead{description}  }
\startdata
\verb!fiducial!  &  The fiducial set of runs: $150 \!\times\! 0.02$-M$_\oplus$ bodies\\[.12cm]
\verb!numdown!   &  Initial number of bodies half that of the fiducial 
 runs: $75 \!\times\! 0.02$-M$_\oplus$ bodies\\[.12cm]
\verb!numup!     &  Initial number of bodies twice that of the fiducial 
 runs: $300\!\times\!0.02$-M$_\oplus$ bodies\\[.12cm]
\verb!massdown!  &  Initial body mass half that of the fiducial 
runs: $150 \!\times\! 0.01$-M$_\oplus$ bodies\\[.12cm]
\verb!massup!    &  Initial body mass twice that of the fiducial 
runs: $150 \!\times\! 0.04$-M$_\oplus$ bodies\\[.12cm]
\verb!notidal!   &  Same as the fiducial set of runs, but with tidal effects ignored\\[.12cm]
\verb!merge!     &  Same as the fiducial set of runs, but all collisions assumed
to be mergers\\
\end{deluxetable}

To understand how changes to the initial number of bodies affect
the outcomes, we ran three sets of simulations with
initial conditions and MFM that matched the three fiducial sets except
that the
initial number of bodies was 75 (instead of 150), as well as three sets  
of simulations matching the 
fiducial sets except with 300 initial bodies.  The sets with 75
initial bodies have names prepended \verb!numdown! and the sets with
300 initial bodies have 
names prepended  \verb!numup!.

To understand how changes to the total mass present in the initial
bodies affected the outcomes, we ran three sets of simulations with
initial conditions and MFM that matched the three fiducial sets except
that the initial body masses were 0.01 M$_\oplus$ (instead of 0.02
M$_\oplus$), as well as three sets of simulations matching the fiducial sets
except with initial
body masses of 0.04 M$_\oplus$.  
The sets with 0.01-M$_\oplus$ initial bodies have names prepended
\verb!massdown! and the sets with 0.04-M$_\oplus$ initial bodies have
names prepended \verb!massup!.

We also ran a few sets of control simulations.  Three sets had identical
initial conditions and MFM as the fiducial sets but the effects of
stellar tides were ignored, i.e., the collision algorithm of C13 was
followed exactly and Equations~(\ref{eq:veschillmod}) and
(\ref{eq:qstartidal}) were ignored (i.e., $v_e$ and $Q^*$ were used instead
of $v_e'$ and $Q^*_\textrm{tidal}$).  These sets have names prepended 
\verb!notidal!.  
A fourth set of control
simulations had perfect mergers for all collisions, i.e., all
collisions were perfectly inelastic and no fragmentation occurred.
The name of this set is prepended  \verb!merge!.

Table~\ref{tab:run_names_table} summarizes the prefixes used in our
naming scheme for the sets of simulations.

\begin{figure}
\includegraphics[width=\columnwidth]{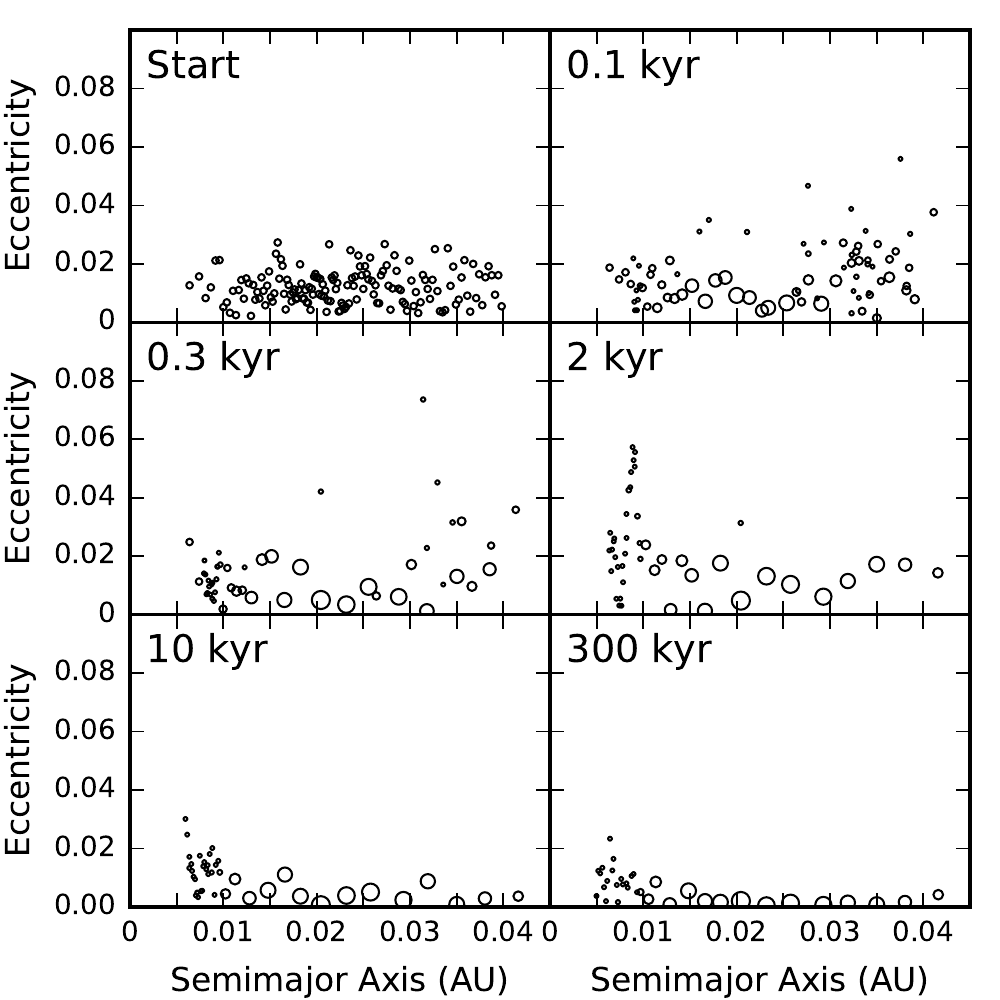}
\caption{\label{fig:fig1}  Eccentricity versus semi-major axis of all
  the bodies for one of the \protect\UseVerb{mfm12} simulations, 
taken at six different
  snapshots.  The symbol radius is proportional to the radius of each
  body.  The minimum fragment mass (MFM) is
  $4.0\!\times\!10^{-3}$ M$_\oplus$, the largest value used in this work.
  At the end of the run, after $3\!\times\!10^5$ years, the system contains
  16 bodies on low-eccentricity orbits outside $r_\mathrm{Roche}\!
  = \! 0.0089$ AU and
  18 fragments inside $r_\mathrm{Roche}$.}
\end{figure}

\begin{figure}
\includegraphics[width=\columnwidth]{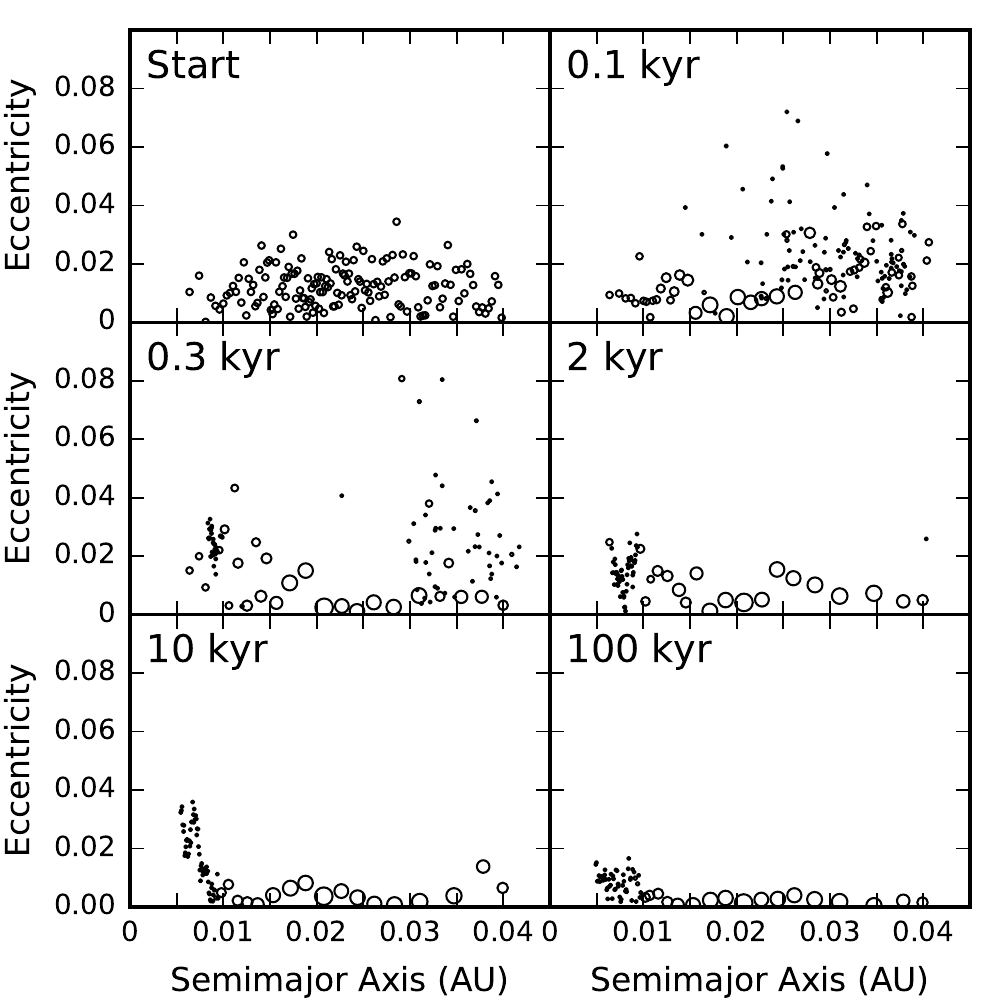}
\caption{\label{fig:fig2} Same as Figure~\ref{fig:fig1}, but for one
of the \protect\UseVerb{mfm04} simulations, with MFM set to
${\sim}1.33\!\times\!10^{-3}$ M$_\oplus$, a factor of three smaller than in
Figure~\ref{fig:fig1}.   Note that the final snapshot is at
100~kyr instead of 300~kyr as in Figure~\ref{fig:fig1}; this
simulation was not run as long due to the
larger computational demands of runs with smaller MFM.  At the
end of the run, the system contains 24 bodies on low-eccentricity
orbits outside $r_\mathrm{Roche}$ (several of which are fragments
inside $1.1$ $r_\mathrm{Roche}$) and 45 fragments inside
$r_\mathrm{Roche}$.}
\end{figure}

We found that ${\sim}$10\% of the runs 
failed to finish.  An investigation showed that
these runs were effectively frozen immediately 
after collisions that produced ${\gtrsim}90$ 
fragments.  We believe that the large number of bodies produced 
in close proximity to each other after these particularly fragmenting
collisions  
have caused these runs to get bogged down in 
the Bulirsch-Stoer portion of the
hybrid-symplectic integration of {\it Mercury}.  We neglect
these failed-to-complete runs in our analysis, but believe that doing
so should not bias our conclusions since the results do
not appear to depend on 
the details of the collision history during the run.  We also ran
supplementary simulations, of which a sufficient number completed to
make up for the runs that failed to complete.  For
supplementary \verb!numup! and \verb!massup! runs, it was necessary to
set a maximum number of fragments that could be produced in a single
collision to allow any of the runs to successfully complete. The
maximum number of fragments used for these runs was 50.


\section{Results}
\label{sec:results}
We present the results from our fiducial runs in
Section~\ref{sec:4.1} and then describe the results from our
variations on the fiducial runs in Section~\ref{sec:4.2}.

\subsection{Fiducial runs}
\label{sec:4.1}

Figure~\ref{fig:fig1}
shows several snapshots of  eccentricity vs. semi-major axis 
from one of the \UseVerb{mfm12} simulations.  
Within 100 years, several of the bodies have already built up to large
fractions of their final masses.  This may seem fast, but by 100 years
bodies with $a\!=\!0.01$ AU have gone through $10^5$
orbits. 
Over the remainder
of the simulation, the planets continue building up to their
final masses by accumulating the remaining smaller bodies and merging with
each other. 
 The fragments inside (and just outside)
  the Roche radius never 
consolidate, as expected from the collision algorithm, and end up
with larger (but still small) eccentricities at the end of the
simulation than the planets outside $r_\mathrm{Roche}$.
Figure~\ref{fig:fig2} is the same as Figure~\ref{fig:fig1} but for one
of the \UseVerb{mfm04} simulations, which has an MFM that is
three times smaller than the run shown in Figure~\ref{fig:fig1}.
Other than a larger buildup of 
fragments inside $r_\mathrm{Roche}$ and an increased number of
 fragments outside $r_\mathrm{Roche}$ in the intermediate stages,
 the simulation proceeds qualitatively the same
as that in Figure~\ref{fig:fig1}.  In both runs, the fragments
are unable to accumulate for $a\!\lesssim\!0.01$ AU, which is
${\sim}10$\% larger than 
$r_\mathrm{Roche}$. This behavior occurs in
all \UseVerb{mfm12}, \UseVerb{mfm08},
and \UseVerb{mfm04} runs.  The lack of planet formation 
between $r_\mathrm{Roche}$ and ${\sim}1.1$~$r_\mathrm{Roche}$ is
presumably due to the large tidal 
forces acting on the bodies in that regime, which
enhance the amount of fragmentation that occurs and makes it more
difficult for bodies to merge, especially for 
collisions between bodies of similar size ($\alpha\!\approx\! 1$).
This behavior has been studied in
detail in the context of ring 
particles by, e.g., \cite{canup1995} and \cite{yasui2014}.
Our control runs (\verb!merge! and \verb!notidal!), which
ignore tidal effects, form planets at all semi-major axes that
  were populated in the initial state, even well inside
$r_\mathrm{Roche}$.

 The eccentricities of the planets in Figure~\ref{fig:fig1} get damped
 over time, and the final 
mean eccentricity of the 16 bodies outside 
$r_\mathrm{Roche}$  is only 0.0028.
To determine the source of the eccentricity damping, 
we re-ran some of the fiducial runs with a single modification: at
 $1.5\!\times\!10^4$ years, all bodies inside a radius $r_{\rm clear}$
 were removed.  We experimented with $r_{\rm clear}\! =\!
 r_\mathrm{Roche}$ and $r_{\rm clear}\! =\!1.1$~$r_\mathrm{Roche}$.
 Eccentricity damping still occurred in the
simulations of the $r_{\rm clear}\! =\! r_\mathrm{Roche}$ sets but the
eccentricity damping in the 
 $r_{\rm clear}\! =\! 1.1$~$r_\mathrm{Roche}$ sets ceased at the time
 of body removal.  Thus, the
eccentricity damping seems to be due mainly to the bodies between 
$r_\mathrm{Roche}$ and $1.1$~$r_\mathrm{Roche}$.

\begin{figure*}
\includegraphics[width=\textwidth]{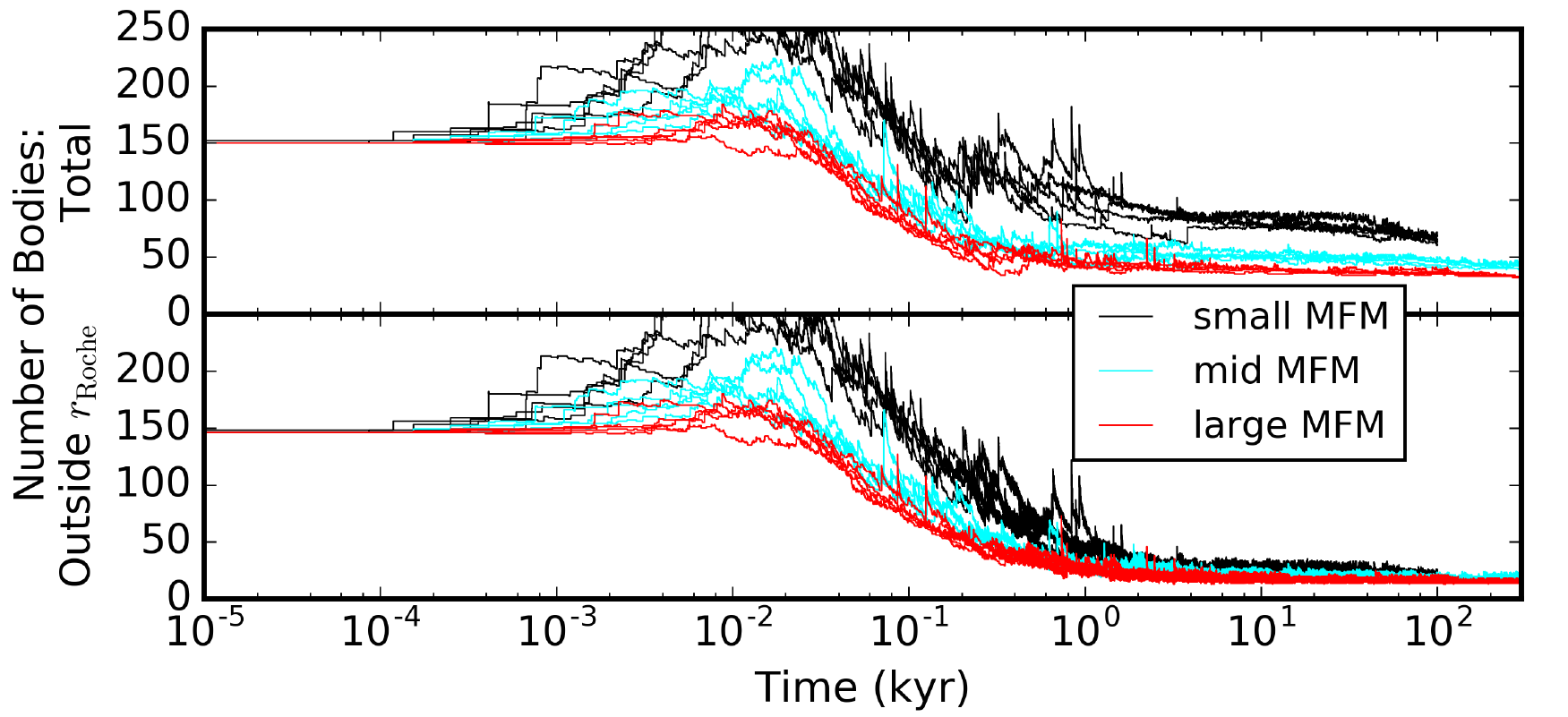}
\caption{\label{fig:fig3} The number of bodies as a function of time
  for the five simulations in each of
  the three fiducial   sets.  The top panel shows the total 
  number of bodies, while the bottom panel shows only those bodies
  that are outside $r_\mathrm{Roche}$.  The simulations from each
 set are encoded by color: ``large MFM'' corresponds
  to \protect\UseVerb{mfm12} ($\mathrm{MFM}\!=\!1.2\!\times\!10^{-8}$~M$_\odot$) and is
  shown in red,
  ``mid MFM'' corresponds 
  to \protect\UseVerb{mfm08} ($\mathrm{MFM}\!=\!0.8\!\times\!10^{-8}$~M$_\odot$) and is
  shown in cyan, and
  ``small MFM'' corresponds 
  to \protect\UseVerb{mfm04} ($\mathrm{MFM}\!=\!0.4\!\times\!10^{-8}$~M$_\odot$) and is
  shown in black.
  All simulations 
  converge on a final set of planets outside $r_\mathrm{Roche}$ with
  similar properties ($N\simeq 15$, $\langle e \rangle \simeq 0.003$).}
\end{figure*}

\begin{figure*}
\includegraphics[width=\textwidth]{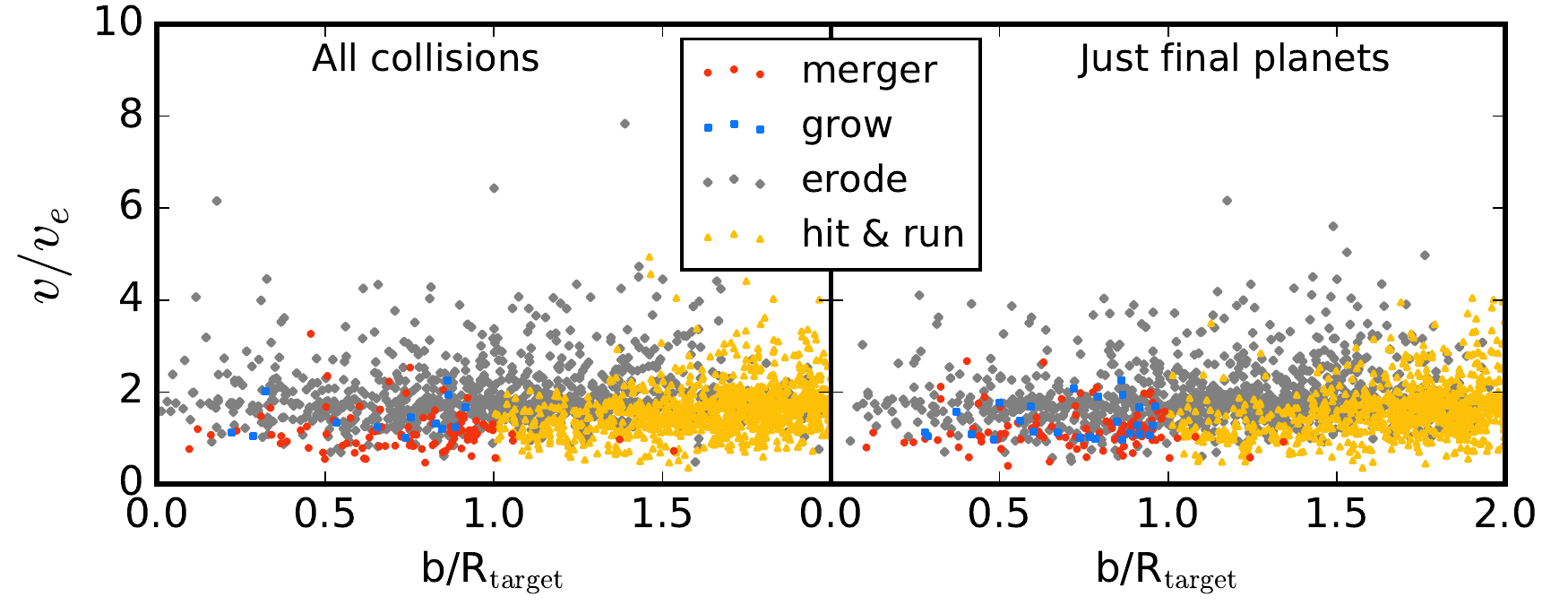}
\caption{\label{fig:fig4} A random sample of collision outcomes for
  the same \protect\UseVerb{mfm12} simulation  shown
  in Figure~\ref{fig:fig1}. The plots show impact velocity versus impact
  parameter for collisions occurring outside
  $r_\mathrm{Roche}$.  The impact parameter $b$ is defined as the
  distance between the centers of the two bodies at the time of the
  collision, projected perpendicular to the relative velocity vector. 
Only a random sample of 2000 collision outcomes is
  shown in each panel to prevent overcrowding.  
The left panel shows a random sampling of all collisions (total
  number: 88,659)
  and the right panel shows a sampling of the collisions involving the bodies that
 become the final planets (total number: 4,625).    
  In the legend, ``merger'' refers to collisions that were 
treated in the code as mergers or as graze-and-merge collisions,
``hit \& run'' refers to collisions classified as hit-and-run (high
  impact parameter, no mass loss for target body), while among
non-hit-and-run fragmenting collisions those with the largest body
losing mass in the collision are labeled as ``erode'' and those
with the largest body coming out of the collision with at least as
much mass as before the collision are labeled as ``grow''.
 The division at $b/R_{\mathrm{target}}\! =\! 1$ between
  ``grow'' collisions and ``hit \& run'' collisions is due to the
  definition of hit-and-run collisions, as detailed in
  LS12 and C13.  There are far more erosive and hit-and-run
  collisions than collisions that lead to mass growth;
  however, a closer look reveals that most these are within ${\sim}1.1$ $r_\mathrm{Roche}$.  See text for details.}
\end{figure*}

\begin{figure*}
\includegraphics[width=\textwidth]{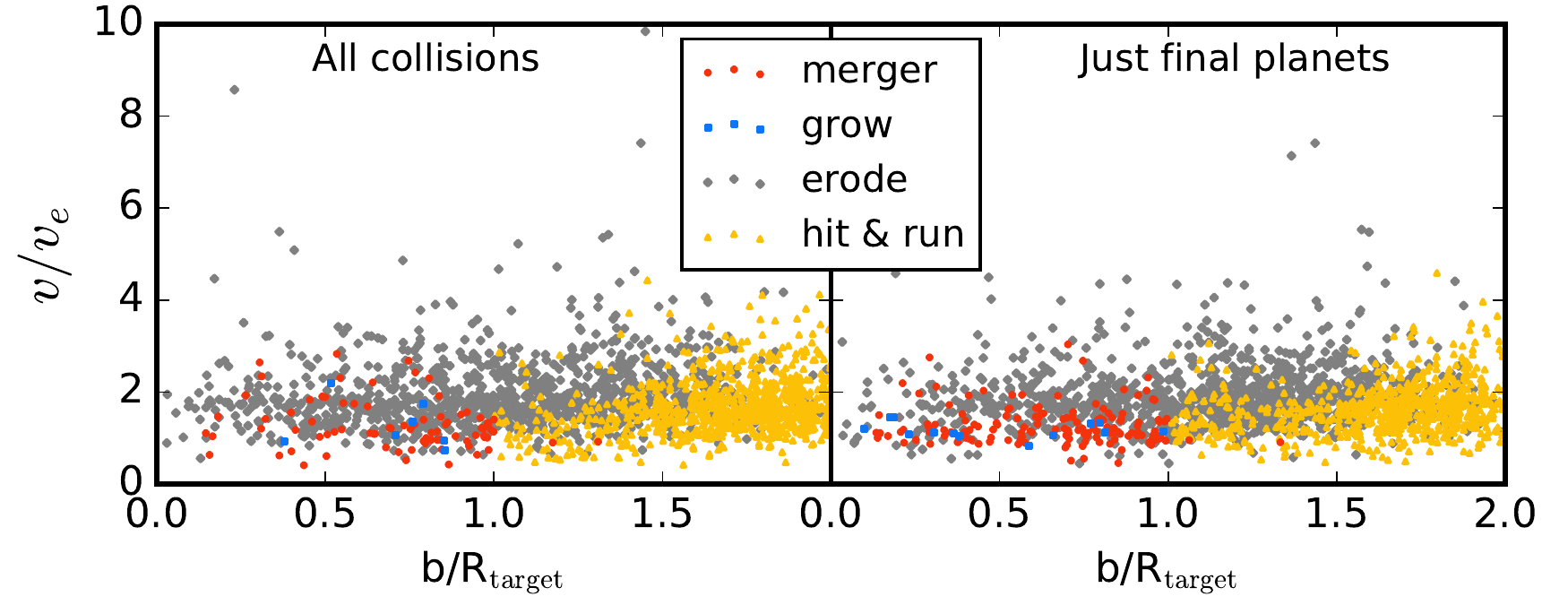}
\caption{\label{fig:fig5} Same as Figure~\ref{fig:fig4}, but for a
  simulation from the \protect\UseVerb{mfm04} set (the same simulation as shown
  in Figure~\ref{fig:fig2}).  The total number of
  collisions is 344,847 and the total number of collisions involving
  the bodies that become the final planets is 10,003.}
\end{figure*}

\begin{figure*}
\includegraphics[width=\textwidth]{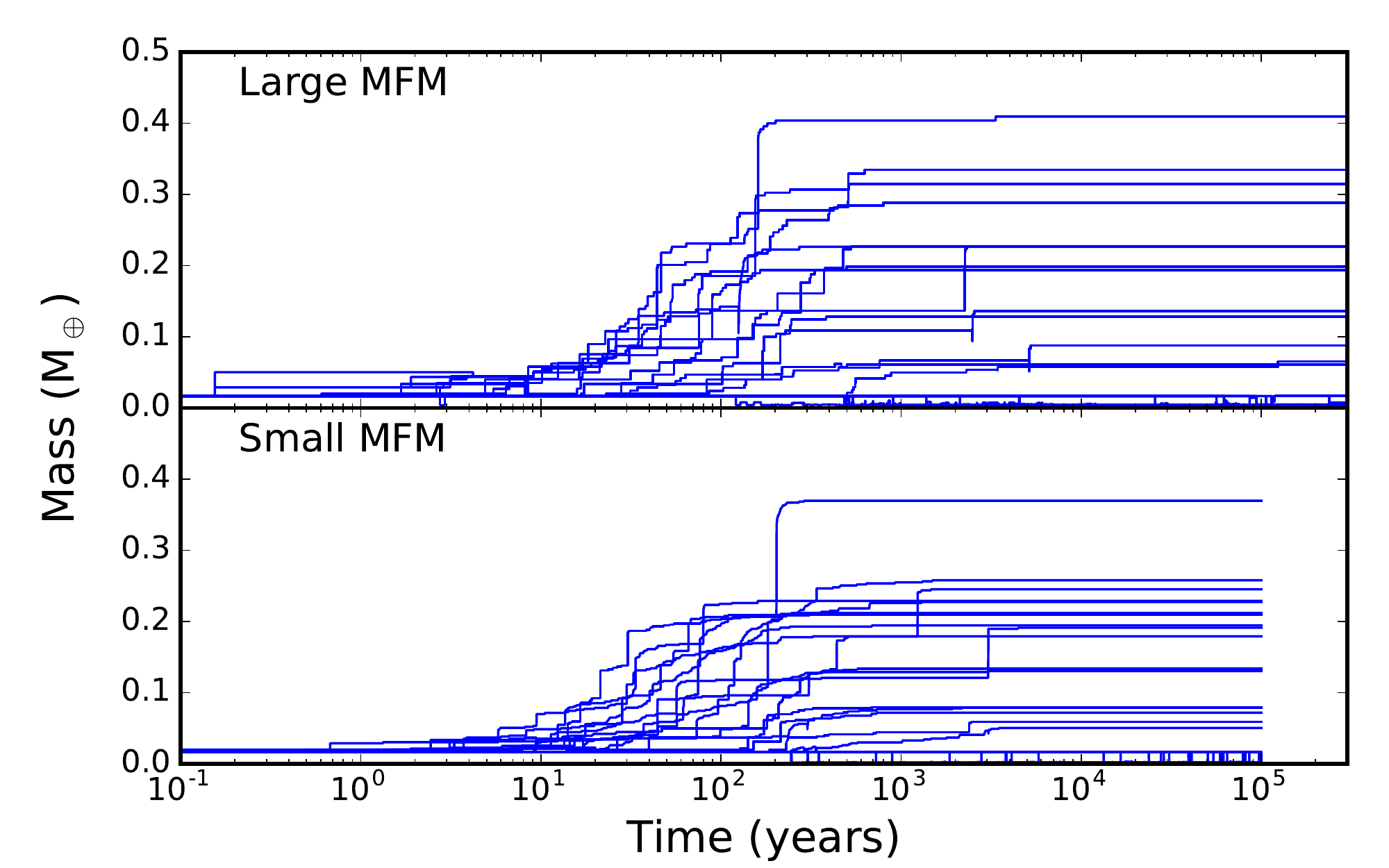}
\caption{\label{fig:fig6} Mass as a function of
  time for bodies that become the final planets.
The top panel shows one of the simulations from
the \protect\UseVerb{mfm12} set (the same simulation as in
Figures~\ref{fig:fig1} and \ref{fig:fig4})
while the bottom panel shows one of the simulations from
the \protect\UseVerb{mfm04} set (the same simulation as in
Figures~\ref{fig:fig2} and \ref{fig:fig5}). 
The final planets outside 1.1 $r_\mathrm{Roche}$ 
exhibit a steady march to their
final masses, with relatively few erosive collisions.  Note that the
simulations of the bottom panel were run for less time than those of
the top panel, as described in Section~\ref{sec:initialconditions}.}
\end{figure*}

Figure~\ref{fig:fig3} shows the number of bodies as a function of time
for all the simulations in
the \UseVerb{mfm12}, \UseVerb{mfm08},
and \UseVerb{mfm04}  sets.  
 The top panel shows the total number of bodies while
the bottom panel shows only those bodies outside $r_\mathrm{Roche}$. 
In the latter case, the number of bodies converges to a
constant value of ${\sim}15$ bodies outside
$r_\mathrm{Roche}$ by ${\sim}100$ kyr, independent of MFM. In what 
follows, we regard the bodies present outside $r_\mathrm{Roche}$
at the end of the integrations as
the ``final planets'' (even though those inside ${\sim}1.1$
$r_\mathrm{Roche}$ have not built up to typical planetary masses), while the
bodies present inside 
$r_\mathrm{Roche}$ at the end of the integrations are referred
to as ``fragments'', because these bodies have fragmented down to 
nearly the MFM. The number of
fragments has also converged to a constant
value in each run, as can be seen in the top panel of Figure~\ref{fig:fig3}.  

In both panels of Figure~\ref{fig:fig3}, the number of bodies  shows a 
persistent trend downward with time in each simulation after the first
30 years or so.
The downward trend is interrupted occasionally by collisions that
produce a large 
number of fragments, especially for lower-MFM runs.  However, these
fragments get quickly
reaccumulated and there is never any runaway
fragmentation. In all of our simulations, a system of 14--24 planets is
always formed outside $r_\mathrm{Roche}$ (with 11--17 bodies outside
$1.1$ $r_\mathrm{Roche}$).

\begin{figure*}
\includegraphics[width=\textwidth]{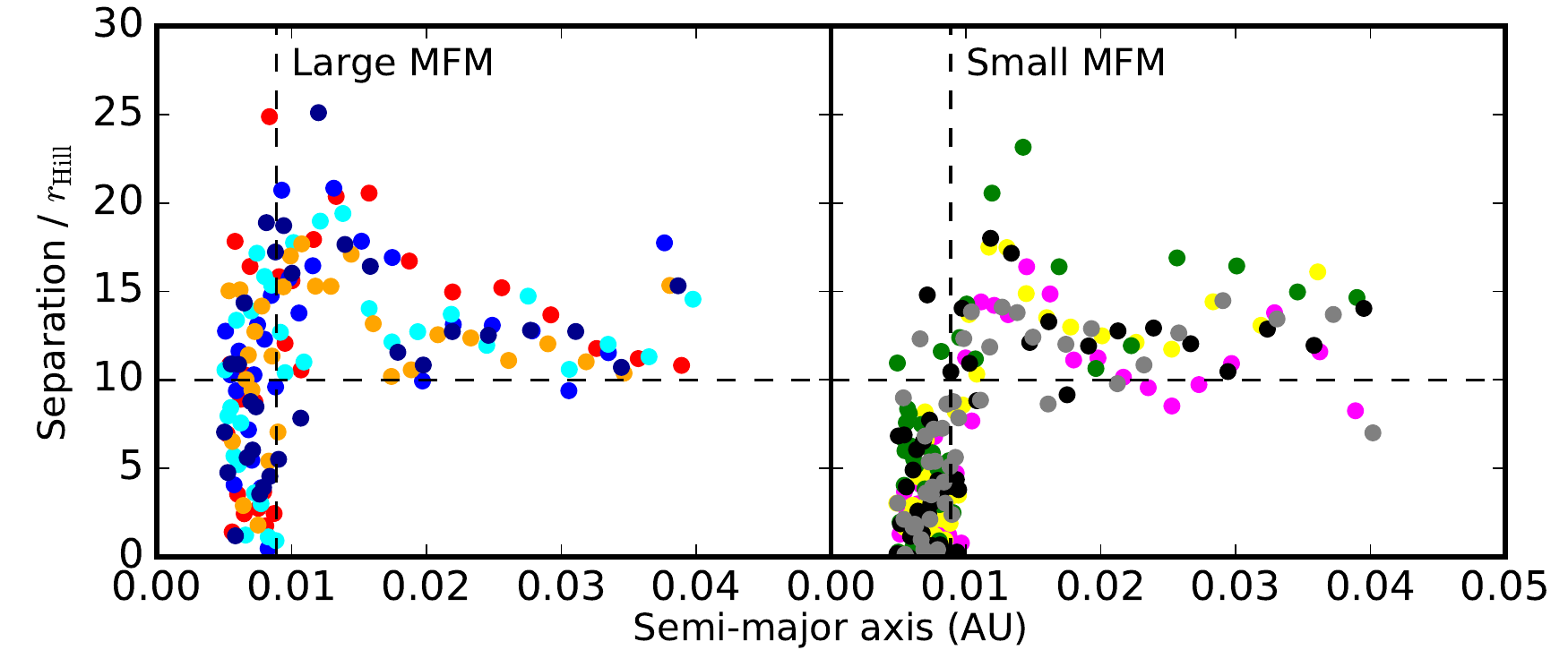}
\caption{\label{fig:fig7} The separation between
  adjacent bodies (defined as the difference between apastron of the
  inner body and periastron of the outer body) in units of 
 $r_\mathrm{Hill}$ as a function of semi-major axis at the end of the
  simulation.  Here, the semi-major axis is calculated as the mean of
  the apastron of the inner body and the periastron of the outer
  body. Each color
  corresponds to bodies from a single simulation.  The vertical
  dashed line shows the location of $r_\mathrm{Roche}$.  The left panel
  shows results from the \protect\UseVerb{mfm12} simulations
  while the right panel shows results from the \protect\UseVerb{mfm04}
  simulations.   The minimum separation required for stability
  over $10^{10}$ orbits (10 million years at
  0.01 AU) is ${\sim}$10 $r_\mathrm{Hill}$
  (\citealt{pu2015}), marked by a horizontal dashed line. The separations of
  bodies inside $r_\mathrm{Roche}$ are smaller, partly because the
  eccentricities are higher and partly because we do not allow these
  bodies to merge to form more stable systems.}
\end{figure*}

The left panel of Figure~\ref{fig:fig4} shows a sample of the 
collision outcomes in one of the \UseVerb{mfm12} simulations and the
right panel shows a sample of the subset 
of collision outcomes involving the bodies that become
the final planets.  Figure~\ref{fig:fig5} shows the same for one of the
\UseVerb{mfm04} runs. 
In these figures, and in the discussion that follows, only collisions
that occurred outside $r_\mathrm{Roche}$ are considered.
The collisions are classified into one of four categories: ``merger'',
``grow'', ``erode'', and ``hit \& run''.  Collisions that were
treated in the code as mergers or as graze-and-merge collisions are here
collectively classified as ``merger'',
hit-and-run collisions are classified as ``hit \& run'', while among
non-hit-and-run fragmenting collisions those where the largest body
loses mass are classified as ``erode'' and those
where the largest body comes out of the collision with at least as
much mass as before are classified as ``grow''.
In the right panel of Figure~\ref{fig:fig4}, 
6\% of collisions 
were classified as ``merger'' and 1\% were classified as ``grow'', compared to
5\%  and 0.9\% for all collisions.
These percentages are characteristic of all of
the \UseVerb{mfm12} runs.
Remarkably, the bodies that became the final planets had a larger percentage of
``erode'' collisions than did all the bodies: 52\% for the run in
Figure~\ref{fig:fig4} for the final planets as
compared with 46\% among all the collisions.  For all
five \UseVerb{mfm12} runs, the percentage of ``erode''
collisions was higher among the bodies that became the final planets
than among all the 
bodies.  Similar patterns were found among
the  \UseVerb{mfm04} runs, one of which is shown in
Figure~\ref{fig:fig5}: all five of the runs had a greater fraction of
``merger''  
and ``grow'' collisions among collisions involving just the bodies
that became the final 
planets than all the bodies, 
and four of the five runs had a greater fraction of ``erode''
collisions among collisions involving just the bodies that became the 
final planets than all the bodies.  For the run shown in
Figure~\ref{fig:fig5}, 7\% of
collisions involving bodies that became the final planets were
``merger'', 0.7\% were ``grow'', 
and 55\% were ``erode'', as compared to 4\% ``merger'', 0.3\%
``grow'', and 54\% ``erode'' collisions among all the bodies.  For
all ten of the runs in the \UseVerb{mfm12}
and \UseVerb{mfm04} sets, the fraction of ``hit \& run''
collisions is lower among collisions involving just the bodies that
became the final planets than among all the bodies. 

It may seem strange that the collisions involving just the bodies that
became the final planets had
a larger fraction of ``erode'' collisions than for all the bodies,
since those bodies were able to gain sufficient mass to become planets by the end
of the simulations.
However, most of these erosive collisions occurred close to
$r_\mathrm{Roche}$, while further out from $r_\mathrm{Roche}$ a larger
fraction of collisions lead to mass growth.  
Close to $r_\mathrm{Roche}$, tidal effects cause a larger fraction of
the collisions to be fragmenting.
Specifically, for the bodies that became the final planets of 
the \UseVerb{mfm12} run shown in Figure~\ref{fig:fig4}, 99.5\%
of ``erode'' 
collisions occurred inside 0.01 AU (${\sim}1.13$ $r_\mathrm{Roche}$)
as well as 86\% of hit \& run collisions, while only 8.8\% of ``merger''
and 3.4\% of ``grow'' collisions occurred inside 0.01 AU.  Thus, the
vast majority of erosive collisions occur close to $r_\mathrm{Roche}$,
where tides are the strongest, and the majority of collisions leading
to mass growth occur outside 0.01 AU, where the final planetary system
forms. Comparable numbers are found for all runs in
the \UseVerb{mfm12}, \UseVerb{mfm08},
and \UseVerb{mfm04} sets.  In short, the bodies outside
${\sim}1.1$ $r_\mathrm{Roche}$ that become the final
planets mostly engage in 
collisions that lead to mass growth, while most of the fragmenting
collisions occur inside
${\sim}1.1$ $r_\mathrm{Roche}$, where tidally enhanced fragmentation
prevents a steady accumulation of mass.

Figure~\ref{fig:fig6} shows the masses of the bodies that become the 
final planets as a function of time.  
In this work and for this figure, the largest remnant in a collision
adopts the name of 
the more massive of the two impacting bodies. 
The final planets outside 1.1 $r_\mathrm{Roche}$ have masses between 
${\sim}0.05$ and ${\sim}0.4$~M$_\oplus$.  Each vertical ``jump'' in
mass represents a collision.
 The bodies that undergo significant mass growth (i.e., the final
 planets outside of ${\sim}1.1$ $r_\mathrm{Roche}$) have
nearly uninterrupted marches towards higher masses.  We
must keep in mind, though, that the bodies in Figure~\ref{fig:fig6},
by virtue of their growth and their surviving to the end, exhibit a
selection bias in which 
kinds of collisions they experienced.
The planets assemble very quickly, with most reaching their
final masses within ${\sim}10^3$ years.  

Figure~\ref{fig:fig7} shows the separation between adjacent bodies at the end
of the simulations in units of $r_\mathrm{Hill}$.  Here,
separation is defined as the difference of
  the apastron of the inner body and the periastron of the outer
  body.  The condition
for long-term stability (over $10^{10}$ orbits, or 10 million years at 0.01 AU)
of a final planetary configuration is that the separations
exceed ${\sim}10$ $r_\mathrm{Hill}$ (\citealt{pu2015}).  For
bodies outside $r_\mathrm{Roche}$ (denoted with the vertical dashed
line), this condition is met, with only a small fraction of the bodies
having separation $<\!10$ $r_\mathrm{Hill}$. This result confirms
  our use of $f_3\!=\!10$ in Section~\ref{sec:analytic}.
Inside $r_\mathrm{Roche}$, the fragments show a large spread in
separations and the 
separations are, on average, smaller than those of the planets outside
$r_\mathrm{Roche}$.  In particular,
separations as small as $r_\mathrm{Hill}\!\approx\! 0$ are seen.  The
small values of these separations are partly due to the
relatively high eccentricities (see Figures~\ref{fig:fig1}
and \ref{fig:fig2}) and partly due to the inability of the bodies
inside $r_\mathrm{Roche}$ to merge and
evolve into a dynamically stable system.

\begin{figure}
\includegraphics[width=\columnwidth]{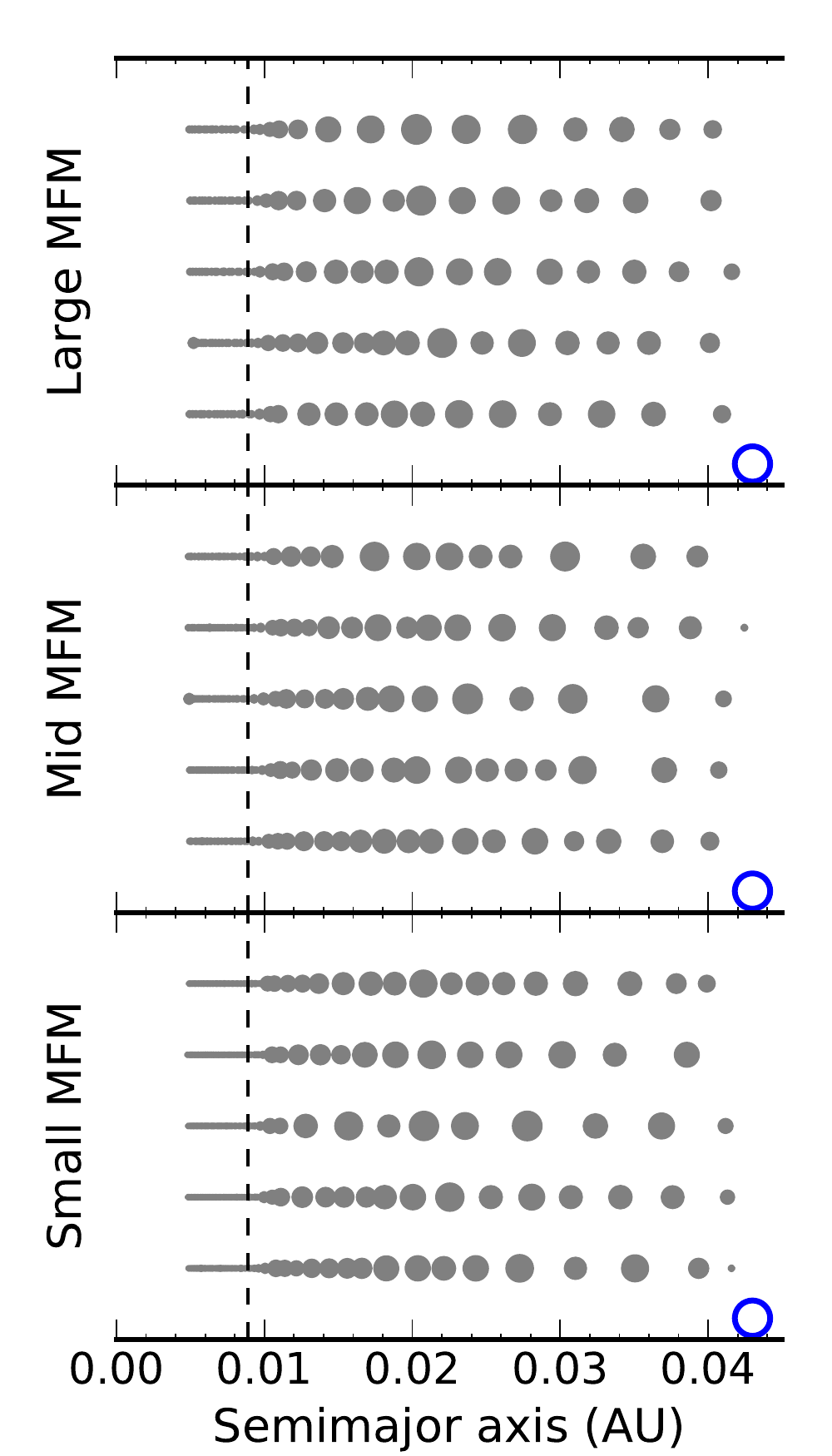}
\caption{\label{fig:fig8} The final planetary configurations for all
  the fiducial sets of simulations.  The top panel shows final
  configurations from the \protect\UseVerb{mfm12} set,
  the middle panel shows final configurations from
  the \protect\UseVerb{mfm08} set,
  and the bottom panel shows final configurations for simulations
  from the \protect\UseVerb{mfm04} set.  Each row of circles corresponds
  to the final bodies of a single simulation.  The radius of
  each circle is scaled by the final radius of the corresponding body,
  assuming the same density for all bodies.  The blue circle in the
  bottom right of each panel shows the radius of a  
  1-M$_\oplus$ body at this density.  The vertical dashed line shows the
  location of $r_\mathrm{Roche}$.  All of our fiducial simulations form
  a stable planetary system at all orbital radii greater than
  ${\sim}1.1$ $r_\mathrm{Roche}$.  The gray horizontal line to the left of ${\sim}$0.01 AU is
  a series of overlapping circles representing the
  closely spaced fragments found at these semi-major axes. }
\end{figure}

Figure~\ref{fig:fig8} shows the final planetary configurations for
our fiducial runs.  The semi-major axes of the  planets with the most
mass in
each system are clustered around ${\sim}0.2 \text{--} 0.3$ AU. As was
noted earlier, tidally enhanced fragmentation prevents planet
formation out to ${\sim}0.01$ AU, which is ${\sim}1.1$
$r_\mathrm{Roche}$.  All of the fiducial runs form a final, stable
planetary system.  The planet-formation efficiency (the fraction of
mass initially present that is incorporated into the final planets) for the mass
initially outside  ${\sim}1.1$ $r_\mathrm{Roche}$ is ${\sim}$100\%,
while inside it is much lower  because the tidally enhanced
fragmentation prevents planets from growing, decreasing with radius to be
approximately zero near $r_\mathrm{Roche}$. 
 The high planet-formation efficiency outside $\sim
1.1$ $r_\mathrm{Roche}$ is partly an artifact of the MFM, which prevents mass loss through collisional
cascades.

\subsection{Other scenarios}
\label{sec:4.2}

\begin{figure}
\includegraphics[width=\columnwidth]{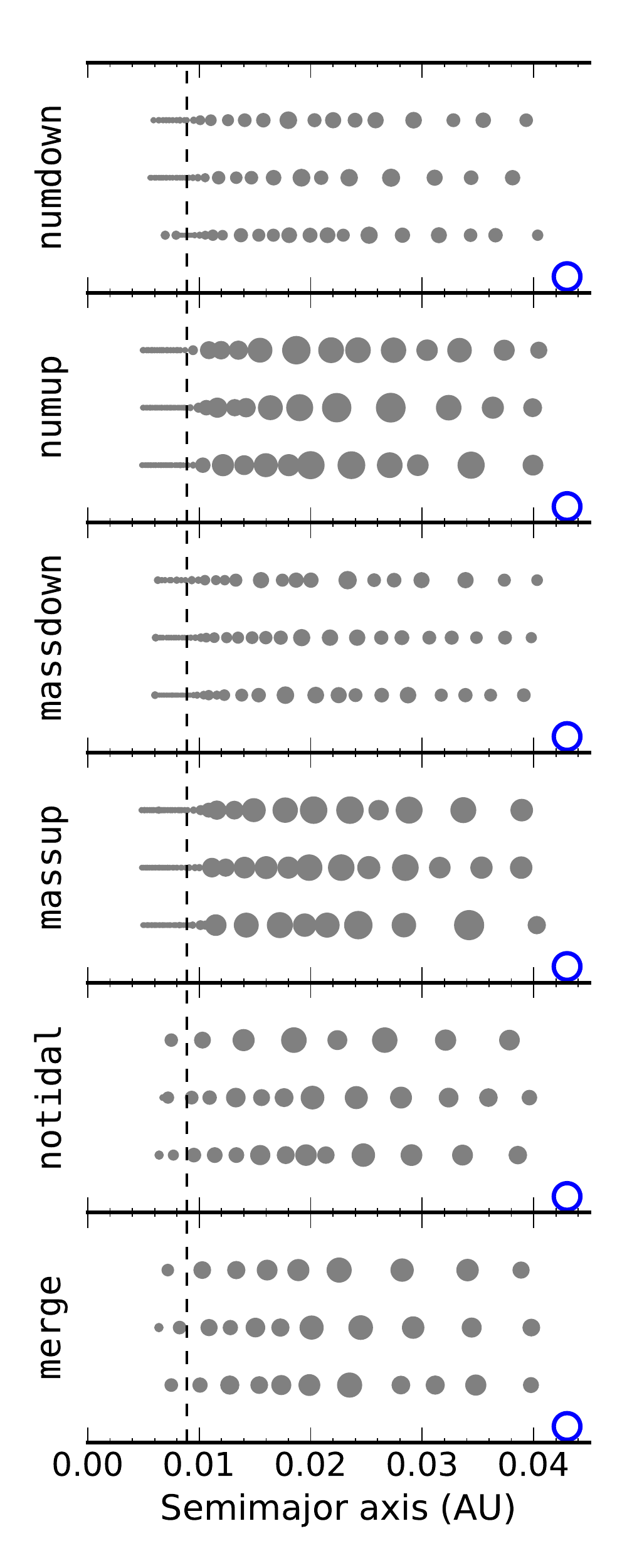}
\caption{\label{fig:fig9} Same as Figure~\ref{fig:fig8} but for final
planetary systems from our initial setups other
than \protect\UseVerb{fiducial}.} 
\end{figure}

Our other sets of runs with tidal effects
(\verb!numdown!, \verb!numup!, \verb!massdown!, \verb!massup!; see
Table~\ref{tab:run_names_table})  also formed planets all the way
down to ${\sim}1.1$ $r_\mathrm{Roche}$.  The distribution and masses
of the planets in the final systems of these sets differed from our
fiducial sets in ways that related to the differing initial
conditions: runs with a larger total initial mass (\verb!numup!
and \verb!massup!) produced a smaller number (${\sim}10$) of
higher-mass planets in their 
final systems than did the fiducial sets, while the runs with a smaller
total initial mass (\verb!numdown! and \verb!massdown!) produced about the
same  number
of planets in their final systems as our fiducial sets, but with lower
masses.  As mentioned previously,
our control runs (\verb!notidal! and \verb!merge!) also form planets
and are able to do so at semi-major axes as small as ${\sim}0.6$
$r_\mathrm{Roche}$ (planet formation at smaller semi-major axes is 
suppressed only because our initial conditions do not have mass interior to
0.005 AU).  The details of how the
various runs got to their final configurations also differed from the
fiducial sets; for instance, more fragmentation occurred in
the \verb!numup! runs than in the fiducial set---likely due to the increased mass available to be turned
into fragments---and the numbers of bodies in the \verb!numup_MFMsmall!
runs was as high as ${\sim}700$ before 
accumulation was able to take over and reduce the total number of
bodies. The planet formation efficiency for these runs for the mass
initially outside  ${\sim}1.1$ $r_\mathrm{Roche}$ is also
${\sim}$100\%, as it was for the fiducial runs.
Figure~\ref{fig:fig9} shows the final configurations of a random
selection of
the \verb!numdown!, \verb!numup!, \verb!massdown!, \verb!massup!,
 \verb!notidal!, and \verb!merge! runs.


\section{Discussion}
\label{sec:discussion}
The analytic work of Section~\ref{sec:analytic} concluded that
collisional fragmentation was not a barrier to planet formation
outside of the Roche radius.  
Our numerical integrations, described in Sections \ref{sec:method} and
\ref{sec:results}, enhanced our analytic work 
 by including the
effects of stellar tides and off-center collisions.
Even with the richer physics, the results of our numerical
integrations agree with the analytic arguments,
except for a small region between 
$r_\mathrm{Roche}$ and ${\sim}1.1$ $r_\mathrm{Roche}$, where
 stellar tides are strong 
enough that even relatively weak collisions lead to fragmentation.

Other works have examined {\it in situ} formation of rocky planets at
small semi-major axes  (\citealp{hansen2012,ogihara2015,dawson2016,moriarty2016,matsumoto2017}).
However, these investigations (i) did not consider 
collisional fragmentation; (ii) examined only initial semi-major axes
larger than 0.04 AU, whereas we focused on smaller semi-major
axes. Our fiducial simulations start with $3M_\oplus$ of solid
material, about three times larger than the mass expected in the minimum-mass solar
nebula between 0.01 AU and 0.04 AU
(\citealp{weidenschilling1977,hayashi1981}), which is comparable
to the disk profiles used by the referenced authors (of course, the
applicability of the minimum-mass solar nebula to these small radii is
highly uncertain). Their simulations typically ran for ${\sim}$1--10
Myr. Our simulations ran for only 0.3 Myr because the dynamical time
at these small radii is much shorter; the shorter integrations
are justified because the numbers and properties of the planets seemed
to be stable well before the end of the integration (see Figure
\ref{fig:fig6}). 

Not considered in this work is the potential loss of mass due to
Poynting--Robertson (PR) drag on dust and small pebbles produced in
collisional cascades.
Bodies of 1-cm radius with density of 1 g cm$^{-3}$ around 
solar-type stars have PR drag
lifetimes of ${\sim}10$~kyr 
at 0.04 AU and ${\sim}1$~kyr at 0.01~AU, with smaller bodies having
shorter lifetimes proportional to their radius
(e.g., \citealt{wyatt1950}). If a sufficient number of other
small bodies are present, lifetimes due to collisional cascades can be even less
than PR lifetimes (e.g., \citealt{wyatt2008}).  Depending on the amount of
small bodies produced in
fragmenting collisions, these processes
may lead to significant mass loss, 
especially near $r_\mathrm{Roche}$ due to the enhanced fragmentation
that occurs there.

Also not considered in this work is the effect of large planets on wider
orbits on the formation and evolution of inner planetary
systems.  A sufficiently close and massive gas giant could 
stir up the bodies and produce larger eccentricities and 
higher collision velocities while the planets are forming.   The
potential effects of gas giants 
depend on the relative times at which they form;
the giant planets would need
to be in place before inner planet formation has ceased in order
to have an effect.  Stirring from giant planets would also increase the 
separation necessary between bodies for long-term stability.

This work assumes no tidal damping of eccentricity or semi-major
axis.  This assumption is justified based on  calculations of the
tidal damping timescales 
(e.g., \citealt{jackson2008}). For these calculations, we assumed 
a Sun-like star, a 0.5-M$_\oplus$ planet with $a\!=\!0.01$ AU, and 
$Q_\mathrm{star}\!=\! 10^{5.5}$ and $Q_\mathrm{planet}\! = \! 100$
for the tidal dissipation parameters of the star and planet.  For these values, 
the eccentricity
damping timescale is ${\sim}10^{5}$~yr and the semi-major axis damping
timescale is ${\sim}10^{10}$~yr, compared to the ${\sim}$100--1000 year
 timescale for planets to form in our
simulations and the $3\!\times\!10^5$ year 
length of our longest simulations.  So, although eccentricity damping is
negligible on the timescales of planet formation, it could affect the
longer-term evolution and cleanup of the system.  Since eccentricity
damping leads to lower-velocity collisions on average, it should not
modify the main conclusion of this work, that planet formation is not
suppressed by fragmentation.

This work also assumes there is no gas left in the disk at the later
stages of planet formation examined here.  With the short ${\sim}100$ year
formation timescale found in our simulations starting from lunar-sized
bodies, though, this assumption may be incorrect.  This, of course,
depends on the gas dissipation timescale at small radii, which is
highly uncertain.

Future work could improve on the collision algorithm of C13 by
implementing a size distribution for the fragments resulting from a
collision, as was done by LS12. This would lead to a richer variety
of mass ratios in collisions than were present in this work. 
Additionally, our initial conditions assumed that all the bodies
started out with 
the same mass.  A more detailed follow-up could look at a more
realistic size distribution for the initial bodies, as well as taking
into account the continued formation of large bodies from accumulating
planetesimals (similar to \citealt{mcneil2005}), which would also lead to a
richer distribution in mass ratios.
  Since, as was shown in Section~\ref{subsec:unequalmass},
collisional growth is robust across all mass ratios, we do not expect
that a more realistic size distribution of collision fragments or
initial bodies will
affect our main conclusions.

As was pointed out in Section~\ref{sec:analytic}, differentiated
bodies behave differently in collisions than homogeneous bodies, and
bodies of the masses encountered in this work ($\gtrsim$lunar mass)
are likely to have significant differentiation.  Head-on collisions
of differentiated bodies 
have higher values of $f_2$ relative to homogeneous bodies 
and thus require higher-velocity
collisions to lead to fragmentation, while off-center collisions have
lower values of $f_2$ (\citealt{asphaug2010}), so these two effects
will partially cancel each 
other.  Even if the off-center collisions with a lower $f_2$ dominate,
the analytic calculation and arguments of Section~\ref{subsec:unequalmass} give
$a_\mathrm{frag}\!\approx\!1.1$ $r_\mathrm{Roche}$  as likely for
equal-mass bodies; 
thus, differentiation of bodies is not expected to  modify
our conclusions significantly. A
collision prescription that combines the general applicability and
scaling laws of LS12 with results of collisions between differentiated
bodies similar to \cite{asphaug2010} would allow for simulating
systems of differentiated bodies in similar style as this work.


\section{Conclusion}
\label{sec:conclusion}
We have carried out both analytic and numerical investigations of
collisional rocky 
planet formation at small semi-major axes (${\sim}0.01$ AU) to
determine whether collisional fragmentation is a barrier to planet
formation. 
Our analytic argument (which ignores tidal effects) predicts that
collisions leading primarily to mass growth are possible all the way
down to the Roche radius and, thus, collisional fragmentation is not a
barrier to planet formation.
Our numerical integrations (which include tidal effects), starting
with ${\sim}$lunar-sized bodies, are able to
form planets all the way down to $a\! \simeq\! 1.1$
$r_\mathrm{Roche}$.  Control integrations
that ignore the effects of tides are able to form planets at even
smaller semi-major axes, consistent with our analytic result.
Our numerical results thus confirm that
collisional fragmentation is not a barrier to rocky planet formation,
except perhaps in a narrow range of distances within 10\% of 
$r_\mathrm{Roche}$. 
The resulting planetary systems are expected to be stable over long time scales
(${\sim}10^{10}$ orbits).

\acknowledgements
We thank Bill Bottke, Eve Lee, and Hilke Schlichting for illuminating
conversations. We also thank the referee for a thorough reading
  of the manuscript and many constructive comments. 
This research has made use of NASA's Astrophysics Data
System Bibliographic Services, the Exoplanet Orbit Database
and the Exoplanet Data Explorer at exoplanets.org (\citealt{han2014}), and
the NASA Exoplanet Archive, which is operated by the California Institute
of Technology, under contract with the National Aeronautics and Space
Administration under the Exoplanet Exploration Program.  The
figures in this work were made using the 
Matplotlib module for Python (\citealt{hunter2007}).

\bibliographystyle{apj}
\bibliography{bibliography}



\end{document}